\RequirePackage[dvipsnames]{xcolor}
\documentclass[journal,twoside,web]{ieeecolor}
\usepackage{tuffc}
\usepackage{cite}
\usepackage{amsmath,amssymb,amsfonts,mathtools}
\usepackage{gensymb}
\usepackage{graphicx}
\usepackage{textcomp}
\usepackage{wrapfig,colortbl}
\makeatletter
\let\NAT@parse\undefined
\makeatother
\usepackage[colorlinks]{hyperref}
\usepackage{xspace}
\usepackage{bm}
\usepackage{cases}
\usepackage{cuted}
\usepackage{booktabs, multirow, multicol}
\usepackage{nicematrix}
\usepackage{threeparttable}
\usepackage{algorithm2e}
\usepackage{pifont}
\usepackage[caption=false]{subfig}
\usepackage{fontawesome5}
\usepackage{etoolbox}
\makeatletter
\@ifundefined{color@begingroup}%
  {\let\color@begingroup\relax
   \let\color@endgroup\relax}{}%
\def\fix@ieeecolor@hbox#1{%
  \hbox{\color@begingroup#1\color@endgroup}}
\patchcmd\@makecaption{\hbox}{\fix@ieeecolor@hbox}{}{\FAILED}
\patchcmd\@makecaption{\hbox}{\fix@ieeecolor@hbox}{}{\FAILED}

\RestyleAlgo{ruled}
\graphicspath{{figures/}}
\DeclareGraphicsExtensions{.pdf,.jpeg,.png}

\newcommand{\ie}{\mbox{i.e.,}\xspace}

\newcommand{\etal}{\mbox{{et al.}}\xspace}
\newcommand{\pinns}{physics-informed neural networks\xspace}
\newcommand{\ivfm}{\textit{i}VFM\xspace}
\newcommand{\nnunet}{\mbox{nnU-Net}\xspace}
\newcommand{\rbpinns}{\mbox{RB-PINNs}\xspace}
\newcommand{\alpinns}{\mbox{AL-PINNs}\xspace}
\newcommand{\invivo}{\mbox{\textit{in vivo}}\xspace}
\newcommand{\cmark}{\text{\ding{52}}}
\newcommand{\xmark}{\text{\ding{56}}}
\definecolor{dark_yellow}{rgb}{1, 0.76, 0}

\hypersetup{
linkcolor=blue,
citecolor=blue,
filecolor=Red,
urlcolor=blue,
menucolor=Red,
runcolor=Red,
linkbordercolor=blue,
citebordercolor=blue,
filebordercolor=Red,
urlbordercolor=blue,
menubordercolor=Red,
runbordercolor=Red
}

\definecolor{abstractbg}{rgb}{1,0.969,0.914}
\def\BibTeX{{\rm B\kern-.05em{\sc i\kern-.025em b}\kern-.08em
    T\kern-.1667em\lower.7ex\hbox{E}\kern-.125emX}}
\begin{document}
\bstctlcite{IEEEexample:BSTcontrol}
\title{Physics-Guided Neural Networks for Intraventricular Vector Flow Mapping}

\author{
Hang Jung~Ling,
Salomé~Bru,
Julia~Puig,
Florian~Vixège,
Simon~Mendez,
Franck~Nicoud,
Pierre-Yves~Courand,
Olivier~Bernard*, and
Damien~Garcia*
\thanks{Manuscript received 19 March 2024; accepted 5 June 2024. This work was supported by: MEGA Doctoral School (ED 162); French National Research Agency (ANR) through the “4D-iVFM” [ANR-21-CE19-0034-01] and “ORCHID” [ANR-22-CE45-0029-01] projects; LABEX PRIMES [ANR-11-LABX-0063] and LABEX CELYA [ANR-10-LABX-0060] of Université de Lyon within the program “Investissements d’Avenir” [ANR-11-IDEX-0007]; GENCI-IDRIS (HPC resources) [2023-AD010313603R1].
}
\thanks{
This work involved human subjects or animals in its research. Approval of all ethical and experimental procedures and protocols was granted by the Human Ethical Review Committee of the University of Montreal Hospital Research Center (CRCHUM).
}
\thanks{
H. J. Ling, J. Puig, P.-Y. Courand, O. Bernard, and D. Garcia are with INSA-Lyon, Universite Claude Bernard Lyon 1, CNRS, Inserm, CREATIS UMR 5220, U1294, F-69621, Lyon, France (e-mail: hangjung.ling@gmail.com, hang-jung.ling@insa-lyon.fr; damien.garcia@creatis.insa-lyon.fr).}
\thanks{
S. Bru, S. Mendez, and F. Nicoud are with IMAG UMR 5149, University of Montpellier, Montpellier, France.
}
\thanks{
F. Vixège is with Univ. Grenoble Alpes, CNRS, Grenoble INP, LEGI, Grenoble, France.
}
\thanks{
P.-Y. Courand is also with the Cardiology Dept., Hôpital Croix-Rousse, Hospices Civils de Lyon, Lyon, France, and the Cardiology Dept., Hôpital Lyon Sud, Hospices Civils de Lyon, Lyon, France.
}
\thanks{
F. Nicoud is also with the Institut Universitaire de France (IUF), France.
}
}

\IEEEtitleabstractindextext{%
\fcolorbox{abstractbg}{abstractbg}{%
\begin{minipage}{\textwidth}\rightskip2em\leftskip\rightskip\bigskip
\begin{wrapfigure}[18]{r}{3in}%
\setlength{\fboxrule}{0.4pt}
\setlength{\fboxsep}{0pt}
\hspace{-3pc}\fcolorbox{subsectioncolor}{white}{\includegraphics[width=2.9in]{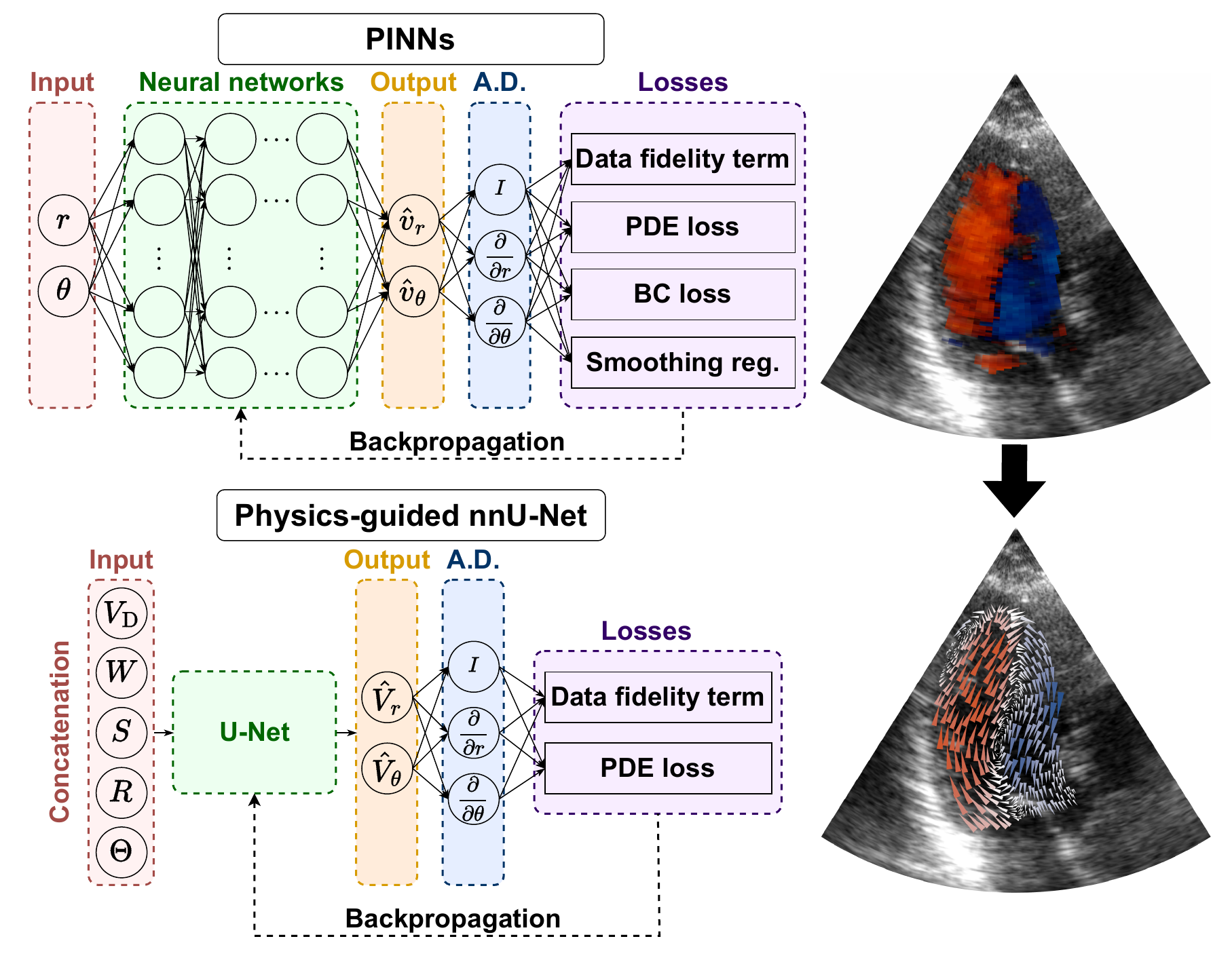}}
\end{wrapfigure}%
\begin{abstract}
Intraventricular vector flow mapping (\ivfm) seeks to enhance and quantify color Doppler in cardiac imaging. In this study, we propose novel alternatives to the traditional \ivfm optimization scheme by utilizing physics-informed neural networks (PINNs) and a physics-guided \nnunet-based supervised approach. When evaluated on simulated color Doppler images derived from a patient-specific computational fluid dynamics model and \invivo Doppler acquisitions, both approaches demonstrate comparable reconstruction performance to the original \ivfm algorithm. The efficiency of PINNs is boosted through dual-stage optimization and pre-optimized weights. On the other hand, the \nnunet method excels in generalizability and real-time capabilities. Notably, \nnunet shows superior robustness on sparse and truncated Doppler data while maintaining independence from explicit boundary conditions. Overall, our results highlight the effectiveness of these methods in reconstructing intraventricular vector blood flow. The study also suggests potential applications of PINNs in ultrafast color Doppler imaging and the incorporation of fluid dynamics equations to derive biomarkers for cardiovascular diseases based on blood flow.
\end{abstract}

\bigskip
\end{minipage}}}

\maketitle

\begin{table*}[!t]
\arrayrulecolor{subsectioncolor}
\setlength{\arrayrulewidth}{1pt}
{\sffamily\bfseries\begin{tabular}{lp{6.75in}}\hline
\rowcolor{abstractbg}\multicolumn{2}{l}{\color{subsectioncolor}{\itshape
Highlights}{\Huge\strut}}\\
\rowcolor{abstractbg}$\bullet$ & Our PINNs, with dual-stage optimization and pre-optimized weights, demonstrated flexibility and performance comparable to the original \ivfm approach for vector flow mapping using color Doppler.\\
\rowcolor{abstractbg}$\bullet${\large\strut} & Our physics-guided \nnunet-based supervised approach achieved robust intraventricular blood flow reconstruction with quasi-real-time inference, even on sparse and truncated Doppler data.\\
\rowcolor{abstractbg}$\bullet${\large\strut} & Our study introduced innovative AI-driven and physics-guided approaches for clinical vector flow mapping, paving the way for enhanced diagnostic accuracy of cardiovascular diseases.\\[2em]\hline
\end{tabular}}
\setlength{\arrayrulewidth}{0.4pt}
\arrayrulecolor{black}
\end{table*}

\section{Introduction}\label{sec:introduction}
\IEEEPARstart{A}{mong} the methods aiming to perform intracardiac flow imaging by color Doppler, intraventricular vector flow mapping (\ivfm) \cite{vixege_physics-constrained_2021, assi_intraventricular_2017,asami_accuracy_2017,meyers_colour-doppler_2020,daae_intraventricular_2021} stands out as a post-processing approach applicable to clinical color Doppler acquisitions. The \ivfm algorithm relies on a constrained least squares optimization scheme (see Section \ref{sec:ivfm} for more details).

Recently, \pinns (PINNs) \cite{raissi_physics-informed_2019} have emerged as a novel approach for data-driven optimization by integrating neural networks (NNs) and the laws of physics during the optimization process. The physical laws, often described by partial differential equations (PDEs), are incorporated into the loss function to enforce the correctness of the solutions. Automatic differentiation \cite{baydin_automatic_2018} has proven to be efficient in computing partial derivatives in PINNs. In cases involving strong non-linear PDEs in the spatiotemporal domain, extensions to PINNs, such as conservative PINNs (cPINNs) and extended PINNs (XPINNs), have been proposed \cite{cai_physics-informed_2021}.

PINNs have found applications predominantly in fluid mechanics \cite{cai_physics-informed_2021}. In the medical field, Arzani \etal \cite{arzani_uncovering_2021} utilized PINNs to recover blood flow from sparse data in 2D stenosis and aneurysm models. Kissas \etal \cite{kissas_machine_2020} applied PINNs to predict arterial blood pressure from 4D flow MRI data. In the ultrasound domain, PINNs have been primarily used for modeling wave propagation \cite{wang_physics-informed_2023}, shear wave elastography \cite{yin_swenet_2023}, and regularizing velocity field given by ultrafast vector flow imaging \cite{guan_towards_2023}.

While PINNs have demonstrated effectiveness on sparse and incomplete data, their application remains unexplored in scenarios where one or more velocity components are missing, as is the case in \ivfm. Color Doppler imaging provides only scalar information—the Doppler velocity, representing the noisy radial velocity—from which we aim to derive both the radial and angular velocity components of intraventricular blood flow.

PINNs often require re-optimization for new cases with different initial or boundary conditions, which can be time-consuming. A potential solution is physics-guided supervised learning \cite{faroughi_physics-guided_2024}, which produces output that adheres to the laws of physics by using a physics-constrained training dataset, with optional physical regularization terms. Once trained, inference can be performed seamlessly on unseen data, provided their distribution closely resembles that of the training dataset.

In this paper, we investigated the feasibility of using physics-based NNs for vector flow mapping, exploring both a physics-guided supervised approach implemented through the \nnunet framework \cite{isensee_nnu-net_2021} and two variants of PINNs. Our contributions included:

\begin{enumerate}
    \item  Training a physics-guided supervised approach based on \nnunet, which showed high robustness on sparse and truncated data with nearly real-time inference speed;
    \item  Implementing two PINNs variants based on the penalty method to perform vector flow mapping, achieving performance comparable to the original \ivfm algorithm;
    \item  Utilizing dual-stage optimization and pre-optimized weights from a selected Doppler frame, which enhanced PINNs' performance and reduced the optimization time of PINNs by up to $3.5$ times.
\end{enumerate}

\begin{figure*}[ht]
    \centering
    \hspace{-1.2em}
    \subfloat[PINNs\label{fig:pinns}]{
        \centering
        \includegraphics[width=0.51\textwidth]{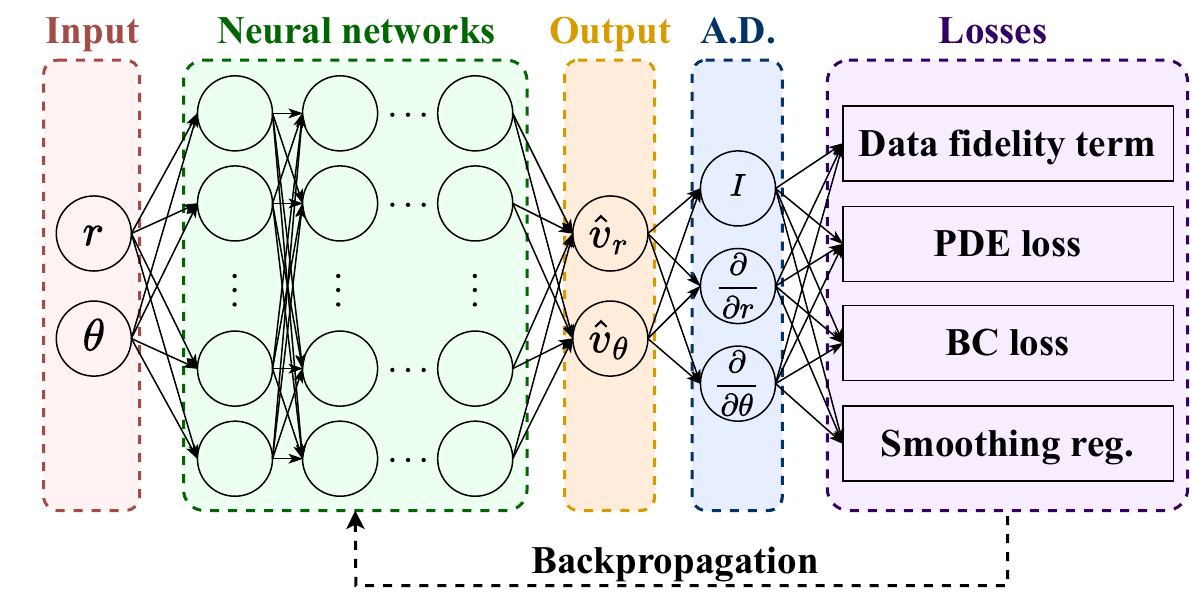}
    }
    \hspace*{\fill}
    \subfloat[Physics-guided \nnunet\label{fig:nnunet}]{
        \centering
        \includegraphics[width=0.47\textwidth]{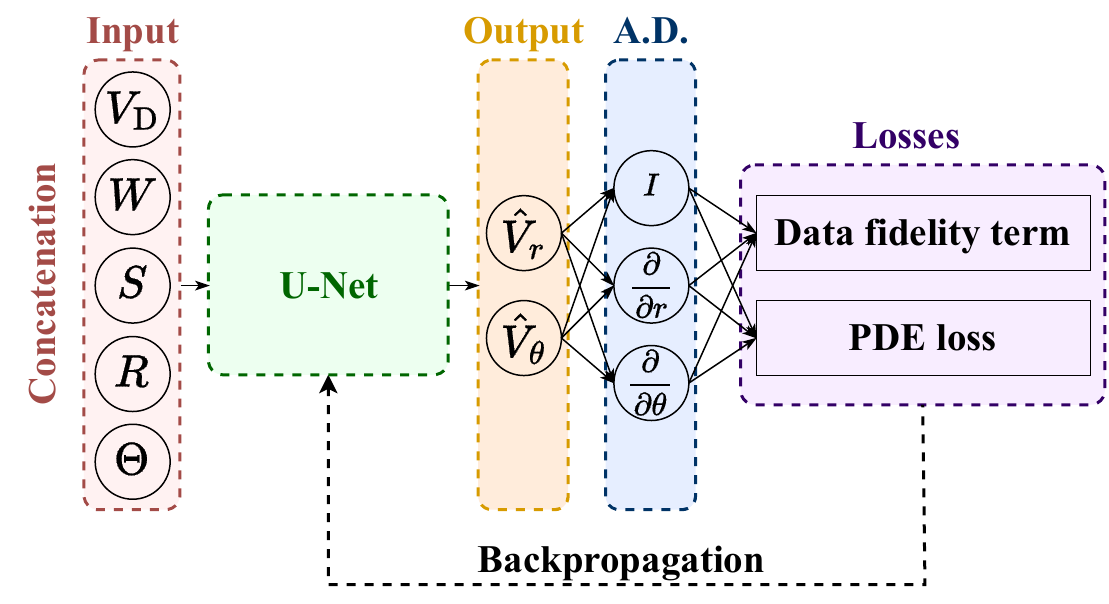}
    }
    \caption{Architectures of PINNs and physics-guided \nnunet. A.D. refers to automatic differentiation. In \protect\subref{fig:pinns}, the 2D input of PINNs has a shape of ($ B \times 2 $), where $ B $ is the batch size. $ r $ and $ \theta $ denote radial and angular coordinates. In \protect\subref{fig:nnunet}, \nnunet takes a 4D input of shape ($ B \times 5 \times 192 \times 40 $), which is the concatenation of sign-inverted dealiased Doppler velocity $ V_\textnormal{D} $, weight matrix $ W $, left ventricular segmentation $ S $, radial coordinate array $ R $, and angular coordinate array $ \Theta $.}
    \label{fig:architecture}
\end{figure*}

\section{Related Work} \label{sec:related_work}
\subsection{Intraventricular Vector Flow Mapping} \label{sec:ivfm}
As introduced in \cite{vixege_physics-constrained_2021}, a vector blood flow map within the left ventricle can be obtained from clinical color Doppler echocardiography by solving a minimization problem. This optimization task is governed by two equality constraints: $ C_1 $, representing the mass conservation equation, and $ C_2 $, representing the free-slip boundary conditions. The first constraint ensures the 2D free-divergence of the optimized velocity field, while the second constraint enforces that the normal component of the blood velocity is zero relative to the endocardial surface. Additionally, a smoothing regularization, further detailed in Section \ref{sec:loss_functions}, is incorporated to impose spatial smoothness of the velocity field. Equation \eqref{eq:math_formulation} expresses the mathematical formulation of this problem.

In \eqref{eq:math_formulation}, $ (\hat{v}_r, \hat{v}_\theta) $ denote the estimated radial and angular blood velocity components. Here, $ \Omega $ stands for the domain of interest, \ie the left ventricle cavity, with its endocardial boundary denoted by $ \partial \Omega $. The term $\omega$ indicates the weights of the data fidelity term. Weights equal to normalized Doppler power values in the range of $ [0, 1] $ were used with \textit{in vivo} Doppler data, as they reflect the reliability of the Doppler velocity. For simulated data, we used $ \omega $ equal to one. $v_\textnormal{D}$ refers to the sign-inverted Doppler velocity (positive velocities for movement away from the probe) to ensure the sign compatibility between $v_\textnormal{D}$ and the $v_r$. The vectors $ \bm{n_\textnormal{W}} = (n_{\textnormal{W}_r},\,n_{\textnormal{W}_{\theta}}) $ and $ \bm{v_\textnormal{W}} = (v_{\textnormal{W}_r},\,v_{\textnormal{W}_{\theta}}) $ represent a unit vector perpendicular to the endocardial wall and a velocity vector of the endocardial wall, respectively.

The \ivfm method \cite{vixege_physics-constrained_2021} linearizes the constrained problem \eqref{eq:math_formulation} and solves it using Lagrange multipliers and a least squares optimization scheme, which reduces the number of supervisedly determined parameters to just one, namely, the smoothing regularization weight.

\begin{equation}\label{eq:math_formulation}
    \begin{gathered}
        \bm{\hat{v}} = (\hat{v}_r, \hat{v}_\theta) = \arg \min_{(v_r, v_\theta)} \underbrace{\int_{\Omega} \omega \left\Vert v_r - v_\textnormal{D} \right\Vert \,d\Omega}_{\mathclap{\text{closely match the Doppler data}}}\\
        \begin{aligned}
            &\text{subject to:}\\
            &\begin{dcases}
            C_1 = r\text{div}(\bm{\hat{v}}) = r\frac{\partial \hat{v}_r}{\partial r} + \hat{v}_r + \frac{\partial \hat{v}_\theta}{\partial \theta} = 0 & \text{on}\, \Omega \\
            C_2 = (\bm{\hat{v}} - \bm{v_\textnormal{W}}) \cdot \bm{n_\textnormal{W}} = (\hat{v}_r - v_{\textnormal{W}_r})n_{\textnormal{W}_r} \\
            \hspace*{2.5em}+ (\hat{v}_\theta - v_{\textnormal{W}_{\theta}})n_{\textnormal{W}_{\theta}} = 0 & \smash{\raisebox{1.5ex}{on $\partial \Omega$}}
            \end{dcases}
        \end{aligned}
    \end{gathered}
\end{equation}

\subsection{Physics-Informed Neural Networks}
Unlike the conventional least squares Lagrangian optimization, such as in \ivfm, PINNs use an iterative scheme with multi-layer perceptrons (MLPs) to iteratively refine and reach the optimal solution. PINNs offer the advantage of being flexible, especially in the optimization scheme, regardless of the linearity of the problem \cite{raissi_physics-informed_2019}. When additional complex physical constraints need to be incorporated, PINNs require minimal architectural modifications, typically requiring only the adaptation of the loss function. However, this process can be challenging when applied to standard approaches involving non-linear constraints.

When addressing a constrained optimization problem using PINNs, the problem is reformulated into a series of loss functions, which often involve conflicting objectives. To manage multi-objective optimization in PINNs, a linear scalarization of the losses is commonly used:
\begin{equation}
    \mathcal{L}_{\mu}(\theta_{\textnormal{NN}}) = \sum_{j=1}^{N}\mu_j \mathcal{L}_j(\theta_{\textnormal{NN}}), \quad \mu_j \in \mathbb{R}_{>0},
\end{equation}
with $ \mu_j $ representing the penalty coefficients or the loss weights, $ \mathcal{L}_{1, \cdots, N} $ being the multiple losses derived from the original constrained optimization problem, and $ \theta_{\textnormal{NN}} $ representing the network parameters. All the losses involved in PINNs' optimization are functions of $ \theta_{\textnormal{NN}} $. However, for better readability of the equations, $ \theta_{\textnormal{NN}} $ is omitted in the subsequent loss expressions.
 
Among the methods with linear scalarization, two notable approaches are the soft constraints and penalty methods. The soft constraints approach uses fixed penalty coefficients throughout the optimization. However, determining optimal coefficients can be challenging, especially as the number of objectives ($ N $) increases, making this approach generally less favored in PINNs optimization.

In contrast to the soft constraints approach, the penalty method involves varying coefficients. Various techniques have been proposed to adapt these coefficients during optimization. Examples include GradNorm \cite{chen_gradnorm_2018}, SoftAdapt \cite{heydari_softadapt_2019} or ReLoBRaLo \cite{bischof_multi-objective_2021}. GradNorm and SoftAdapt dynamically adjust the loss weights based on the relative training rates of different losses. ReLoBRaLo can be seen as a combination of the former two techniques, which incorporates a moving average for loss weights and a random look-back mechanism. The random look-back mechanism is controlled by a variable that determines whether the loss statistics of the previous steps or those of the first step are used to compute the coefficients.

Recently, an alternative method called Augmented Lagrangian (AL) has been proposed for solving constrained optimization problems using PINNs \cite{lu_physics-informed_2021, son_enhanced_2023}. Similar to the penalty method, the AL approach involves penalty terms, but it also introduces a term designed to mimic a Lagrange multiplier.

In situations where optimizing initial or boundary conditions is difficult, some studies have proposed imposing those conditions as hard constraints by utilizing a distance function and an analytical approximation of the conditions \cite{berg_unified_2018, sukumar_exact_2022}. Although it is feasible to impose hard constraints on the output of NNs, it is often challenging and more appropriate for problems with several initial or boundary conditions.

\section{Methods} \label{sec:methodology}
In this study, we addressed the constrained optimization problem of \ivfm \eqref{eq:math_formulation} through NNs aided by physics: a physics-guided supervised approach based on \nnunet and PINNs using the penalty method. Specifically, we studied two variants of PINNs: 1) PINNs with the ReLoBRaLo weight-adapting strategy (\rbpinns); 2) Augmented Lagrangian PINNs (\alpinns). Schematic representations of the general architectures of PINNs and \nnunet for intraventricular vector flow reconstruction are shown in Fig. \ref{fig:pinns} and Fig. \ref{fig:nnunet}, respectively.

The following subsections introduce the loss functions to be optimized in PINNs (Section \ref{sec:loss_functions}), provide implementation details for PINNs (Section \ref{sec:rbpinns} to \ref{sec:pinns_configs}), discuss the physics-guided \nnunet approach (Section \ref{sec:supervised}), and present the evaluation metrics (Section \ref{sec:metrics}).

\subsection{PINNs' Loss Functions}\label{sec:loss_functions}
In line with the previous \ivfm method,  we decomposed the mathematical formulation in  \eqref{eq:math_formulation} into several objectives to be optimized [see \eqref{eq:l1}–\eqref{eq:l4}]: 1) $ \mathcal{L}_1$: data fidelity term; 2) $ \mathcal{L}_2$: mass conservation residual loss (PDE loss); 3) $ \mathcal{L}_3$: boundary condition residual loss (BC loss); 4) $ \mathcal{L}_4$: smoothing regularization.

For the PDE loss, namely $ \mathcal{L}_2 $, the partial derivatives were computed using automatic differentiation . On the contrary, for $ \mathcal{L}_4 $, the partial derivatives were obtained using finite difference methods with 2D convolution kernels, as spatial smoothness could not be computed with automatic differentiation. This was done by setting the weights of the $ 3 \times 3 $ convolution kernels to the central finite difference coefficients with second-order accuracy.

The norm $ \| \cdot \| $ used for computing the $ \mathcal{L}_1 $, $ \mathcal{L}_2 $, and $ \mathcal{L}_3 $ was the Smooth L1 loss or Huber loss with $ \beta=1.0 $ and \textit{sum} reduction over all the samples unless otherwise stated. The Smooth L1 loss uses a squared term if the absolute error falls below $ \beta $ and an absolute term otherwise, making it less sensitive to outliers than the mean squared error.

\begin{numcases}{}
    \mathcal{L}_1 = \omega \left\Vert \hat{v}_r - v_\textnormal{D} \right\Vert &$\, \text{on} \, \Omega$ \label{eq:l1}
    \\
    \mathcal{L}_2 =  \left\Vert r\frac{\partial \hat{v}_r}{\partial r} + \hat{v}_r + \frac{\partial \hat{v}_\theta}{\partial \theta} \right\Vert &$\, \text{on} \, \Omega$ \label{eq:l2}
    \\
    \mathcal{L}_3 = \left\Vert (\hat{v}_r - v_{\textnormal{W}_r})n_{\textnormal{W}_r} + (\hat{v}_\theta - v_{\textnormal{W}_{\theta}})n_{\textnormal{W}_{\theta}} \right\Vert &$\, \text{on} \, \partial \Omega$ \label{eq:l3}
    \\
    \!\begin{aligned}
    \mathcal{L}_4 &= \sum_{k \in \{r, \theta\}}\Biggl\{\left(r^2 \frac{\partial^2 v_k}{\partial r^2}\right)^2 + 2\left(r \frac{\partial^2 v_k}{\partial r \partial \theta}\right)^2 \\
    &\hspace*{1.5em} + \left(\frac{\partial^2 v_k}{\partial \theta^2}\right)^2\Biggl\}
    \end{aligned} &$\, \text{on} \, \Omega$ \label{eq:l4}
\end{numcases}

\subsection{\rbpinns}\label{sec:rbpinns}
\subsubsection{Global Loss}
The global loss function to be optimized in \rbpinns was defined as:
\begin{equation}\label{eq:ReLoBRaLo_loss}
     \mathcal{L}_{\mu, \theta_{\textnormal{NN}}} = \underbrace{\mu_1\mathcal{L}_1}_{\mathclap{\substack{\text{data fidelity} \\ \text{term}}}} + \underbrace{\mu_2 \mathcal{L}_2 + \mu_3 \mathcal{L}_3}_{\mathclap{\text{PDE \& BC losses}}} + \underbrace{\mu_4 \mathcal{L}_4,}_{\mathclap{\substack{\text{smoothing} \\ \text{reg.}}}}
\end{equation}
where $ \mu_1, \mu_2, \mu_3 \in \mathbb{R}_{>0}$ are adaptive penalty coefficients, and $ \mu_4 $ is the smoothing regularization weight. We heuristically set $ \mu_4 $ to $ 10^{-7.5} $.

\begin{algorithm}
    \DontPrintSemicolon
    Initialize $ \mu_j(1)=1, j \in \{1,2,3\}. $\;
    \For {$ i=1,\dots,I $}{
        Forward pass and compute losses\;
        $ \mathcal{L}_j(i) \leftarrow  \mathcal{L}_j $\;
        \If{$ i >= 2 $}{
            $ \hat{\mu}_j^{(i, i-1)} \leftarrow n_{\textnormal{loss}} \times \textnormal{Softmax}\Bigl(\frac{\mathcal{L}_j(i)}{\mathcal{T} \mathcal{L}_j(i-1) + \epsilon}\Bigr) $\; 
            $ \hat{\mu}_j^{(i, 1)} \leftarrow n_{\textnormal{loss}} \times \textnormal{Softmax}\Bigl(\frac{\mathcal{L}_j(i)}{\mathcal{T} \mathcal{L}_j(1) + \epsilon}\Bigr) $\;
            $ \mu_j(i) \leftarrow \alpha \Bigl( \rho \mu_j(i - 1) + (1 - \rho) \hat{\mu}_j^{(i, 1)} \Bigr)$ \\
            \hspace*{1.5em}$+ ( 1 - \alpha ) \hat{\mu}_j^{(i, i-1)} $\;
        }
        Compute final loss using \eqref{eq:ReLoBRaLo_loss} and do backpropagation to update network parameters:
        $ \bm{\theta_{\textnormal{NN}}} \leftarrow \bm{\theta_{\textnormal{NN}}} - \eta_{\theta_{\textnormal{NN}}} \nabla_{\theta_{\textnormal{NN}}} \mathcal{L}_{\mu, \theta_{\textnormal{NN}}}(i) $\;
    }
    \caption{ReLoBRaLo update strategy\label{algo:ReLoBRaLo}}
\end{algorithm}

\subsubsection{Update Strategy for Loss Weights}
Algorithm \ref{algo:ReLoBRaLo} details the ReLoBRaLo update strategy introduced in \cite{bischof_multi-objective_2021} for determining the loss weights, i.e., $ \mu_1 $, $ \mu_2 $, and $ \mu_3 $. In this algorithm, $ n_{\textnormal{loss}} $ represents the total number of losses for which the loss weights are updated; in our case, $ n_{\textnormal{loss}}=3 $. $ \hat{\mu}_j^{(i, i')} $ computes the scaling based on the relative improvement of $ \mathcal{L}_j $ between the iterations $ i' $ and $ i $. $ \mu_j(i) $ is defined as the weight for $ \mathcal{L}_j $ at the $ i^{th} $ iteration, obtained through an exponential decay. The algorithm's hyperparameters consist of $ \alpha $ for the exponential decay rate, $ \rho $ for a Bernoulli random variable with an expected value close to 1, $ \mathcal{T} $ for temperature, and $ I $ for the total number of iterations. We heuristically set $ \alpha = 0.999$, $ \mathbb{E}(\rho) = 0.999$, and $ \mathcal{T} = 1.0$, as this combination yielded the best results for our problem. $ \theta_{\textnormal{NN}} $ denotes the learnable network parameters, $ \eta_{\theta_{\textnormal{NN}}} $ is the learning rate used for updating network parameters, and $ \nabla_{\theta_{\textnormal{NN}}} $ is the gradient of the final loss with respect to $ \theta_{\textnormal{NN}} $.

\subsection{\alpinns}
\subsubsection{Global Loss}
For \alpinns, we defined its global loss as:
\begin{equation}\label{eq:AL_loss}
    \begin{aligned}
     \mathcal{L}_{\lambda, \mu, \theta_{\textnormal{NN}}} &= \underbrace{\mathcal{L}_1}_{\mathclap{\substack{\text{data fidelity} \\ \text{term}}}} + \underbrace{ \bigl \langle \bm{\lambda_1}, \bm{C_1} \bigr \rangle + \bigl \langle \bm{\lambda_2}, \bm{C_2} \bigr \rangle}_{\mathclap{\substack{\text{PDE \& BC losses} \\ \text{(Lagrange multipliers)}}}} \\
     &\hspace*{1.5em}+ \underbrace{0.5 \times \mu \times (\mathcal{L}_2 + \mathcal{L}_3)}_{\substack{\text{PDE \& BC losses (penalty)}}} + \underbrace{\mu_4 \mathcal{L}_4.}_{\mathclap{\substack{\text{smoothing} \\ \text{reg.}}}}
    \end{aligned}
\end{equation}
In this equation, $ \lambda_1 $ and $ \lambda_2 $ are learnable real Lagrange multipliers related to the two constraints: the mass conservation, $ C_1 $, and the free-slip boundary condition, $ C_2 $. The notation $ \bigl \langle \cdot, \cdot \bigr \rangle $ refers to the inner product of two vectors. The learnable penalty coefficient for the two physical constraints is denoted by $ \mu \in \mathbb{R}_{>0} $. Similar to \rbpinns, $ \mu_4 $ was set heuristically to $ 10^{-7.5} $.

\begin{algorithm}
    \DontPrintSemicolon
     Initialize $ \bm{\lambda_j} = \vec{0}, \mu=2, j \in \{1,2\}. $\;
    \For {$i=1,\dots,I$}{
        Forward pass and compute losses\;
        $ \mathcal{L}_{\lambda, \mu, \theta_{\textnormal{NN}}}(i) \leftarrow  \mathcal{L}_{\lambda, \mu, \theta_{\textnormal{NN}}} $\;
        Compute final loss using \eqref{eq:AL_loss} and do backpropagation to simultaneously update network parameters, learnable $ \bm{\lambda_j} $ as well as $ \mu $:\;
        \hspace*{0.5em}$ 
        \begin{dcases}
            \bm{\theta_{\textnormal{NN}}} \leftarrow \bm{\theta_{\textnormal{NN}}} - \eta_{\theta_{\textnormal{NN}}} \nabla_{\theta_{\textnormal{NN}}} \mathcal{L}_{\lambda, \mu, \theta_{\textnormal{NN}}}(i)\\
            \bm{\lambda_j} \leftarrow \bm{\lambda_j} + \eta_{\lambda} \nabla_{\lambda_j} \mathcal{L}_{\lambda, \mu, \theta_{\textnormal{NN}}}(i)\\
            \mu \leftarrow \mu + \eta_{\mu} \nabla_{\mu} \mathcal{L}_{\lambda, \mu, \theta_{\textnormal{NN}}}(i)
        \end{dcases}
        $\;
    }
    \caption{Augmented Lagrangian update strategy} \label{algo:AL}
\end{algorithm}

\subsubsection{Update Strategy for Loss Weights}
We applied the gradient ascent method \cite{son_enhanced_2023} to update $ \lambda_1 $, $ \lambda_2 $, and $ \mu $, as decribed in Algorithm \ref{algo:AL}. In the original approach \cite{son_enhanced_2023}, $ \mu $ remains constant throughout the optimization process, but we proposed to update this penalty coefficient to better adhere to the original AL method \cite{hestenes_multiplier_1969}. In our experiments, the gradient ascent method showed higher optimization stability and was less prone to gradient explosion compared to the original AL update rule proposed in \cite{hestenes_multiplier_1969}. In Algorithm \ref{algo:AL}, $ \theta_{\textnormal{NN}} $, $ \eta_{\theta_{\textnormal{NN}}} $, and $ \nabla_{\theta_{\textnormal{NN}}} $ represents the learnable network parameters, the learning rate for updating network parameters, and the gradient of the final loss with respect to $ \theta_{\textnormal{NN}} $, respectively. $ \nabla_{\lambda_j} \cdot $ and $ \nabla_{\mu} \cdot $ denote the gradient of the final loss with respect to $ \lambda_j $ and $ \mu $; $ I $ indicates the total number of iterations; $ \eta_{\lambda} $ and $ \eta_{\mu} $ are the learning rates for the learnable Lagrange multipliers $ \lambda_j $ and $ \mu $. The selection of appropriate learning rates is critical in preventing gradient overflow when dealing with physical losses that involve unbounded Lagrange multipliers. Their values are discussed in Section \ref{sec:training_strategies_pinns}.

\subsection{Dual-Stage Optimization} \label{sec:dual_stage}
To improve the convergence of our PINNs, we introduced a dual-stage optimization strategy: 1) an optimization stage using the AdamW \cite{loshchilov_decoupled_2018} optimizer for the first 90\% of the iterations to converge to a rough solution; 2) a fine-tuning stage using the L-BFGS \cite{liu_limited_1989} optimizer, which is not sensitive to learning rates, for the remaining iterations to obtain an optimal final solution. This approach aimed to reduce the optimization time of PINNs. An ablation study was performed using \rbpinns to assess the potential improvement of this strategy.

\subsection{PINNs' Architecture, Weight Initialization, and Sampling Strategy}\label{sec:pinns_configs}
\subsubsection{Network Architecture}
For both PINNs implemented in this paper, we utilized an MLP with six hidden layers, each containing 60 neurons, with the \textit{tanh} activation function. This architecture resulted in approximately 18.6k trainable parameters.

\subsubsection{Weight Initialization}
We first applied the dual-stage optimization to a Doppler frame selected at the end of the early filling phase using \rbpinns. The resulting weights were then saved as pre-optimized weights and used as initialization for all subsequent PINNs models before optimization on new Doppler data. This initialization technique aimed to accelerate the optimization process of our PINNs and enhance their performance. A second ablation study was carried out to justify this choice.

\subsubsection{Sampling Strategy}
Leveraging the regularly spaced polar grid of color Doppler imaging and its relatively small size, we utilized all sample points within the left ventricle on the grid for both data and collocation points. These data points were used to compute the data fidelity term, while the PDE loss and smoothness term were evaluated from the collocation points. For the boundary condition residual loss, all extracted points on the boundary were considered.

\subsection{Physics-Guided \nnunet}\label{sec:supervised}
We trained a physics-guided \nnunet (refer to Fig. \ref{fig:nnunet} for its architecture) with configurations similar to those described in \cite[Table I]{ling_dealiasing_2023} on both simulations and \invivo data. We adapted the loss function to L1 loss for supervised regression. Additionally, to enforce mass conservation in the predicted velocity field, we incorporated $ \mathcal{L}_2 $ in the loss function as a physical regularization term with a weight of $ \gamma = 10^{-3} $. Both supervised and regularization terms were masked with the left ventricular binary segmentation, restricting loss computation to the region of interest. The final loss was expressed as:
\begin{equation}\label{eq:supervised_loss}
\begin{aligned}
    \mathcal{L}_{\gamma, \theta_{\textnormal{NN}}} 
    &= \underbrace{\left\Vert \hat{V}_r - V_{r_{\text{ref}}} \right\Vert_1 + \left\Vert \hat{V}_{\theta} - V_{{\theta}_{\text{ref}}} \right\Vert_1}_{\mathclap{\text{data fidelity term}}} \\
    &\hspace*{1.5em}+ \gamma \underbrace{\left\Vert r\frac{\partial \hat{V}_r}{\partial r} + \hat{V}_r + \frac{\partial \hat{V}_\theta}{\partial \theta} \right\Vert_1}_{\mathclap{\text{PDE loss}}},
\end{aligned}
\end{equation}
where $ (V_{r_{\text{ref}}}, V_{{\theta}_{\text{ref}}}) $ represent the reference velocity field given by simulations or predicted velocity field by \ivfm \cite{vixege_physics-constrained_2021}, considered as the gold standard for \invivo data.

Unlike PINNs that directly take coordinates $ (r, \theta) $ as input, \nnunet requires image data. Our \nnunet's input was a concatenation of: 1) dealiased color Doppler image before scan-conversion for \invivo data or alias-free image for simulated data; 2) a weight matrix with normalized Doppler powers in the range of $ [0, 1] $ for \invivo data or containing ones for simulated data; 3) binary segmentation of the left ventricle cavity; 4) radial coordinate array; 5) angular coordinate array. More details about the training dataset are given in Section \ref{sec:train_dataset}.

During training, we applied data augmentations, including random rotation ($ [-15,15]\degree $), random zoom ($ [0.7,1.4] $), and random scanline masking. With the latter, a block of $ n $ consecutive scanlines was randomly masked out with a step size of $ m $. In our experiments, $ m = 10 $ and $ n $ was a random integer between $ 0 $ and $ 9 $. This strategy simulated sparse Doppler data, enhancing the model's robustness and generalizability.

\subsection{Evaluation Metrics}\label{sec:metrics}
We assessed the performance of \rbpinns, \alpinns, and \nnunet using simulated Doppler images from a patient-specific computational fluid dynamics (CFD) model (see Section \ref{sec:CFD_dataset}). Evaluation metrics, including squared correlation ($ r^2 $) and normalized root-mean-square error (nRMSE), were computed by comparing predicted and ground truth velocity fields within the left ventricle.

\subsubsection{Squared Correlation}
We defined the squared correlation as follows:
\begin{equation}
    r^2_{v_k} = \text{Corr}(\hat{v}_k, v_{k_{\text{CFD}}})^2, \quad k \in \{r, \theta\},
\end{equation}
where $ \text{Corr} $ is the Pearson correlation coefficient.

\subsubsection{nRMSE}
For both the radial and angular components, we computed the root-mean-square errors normalized by the maximum velocity defined by:
\begin{equation}
    \text{nRMSE} = \frac{1}{\textnormal{max} \left\Vert v_{\textnormal{CFD}} \right\Vert_2} \sqrt{\frac{1}{n}\sum_{k=1}^n \left\Vert \hat{v}_k - v_{\textnormal{CFD}_{k}} \right\Vert_2^2},
\end{equation}
where $ n $ stands for the number of velocity samples in the left ventricular cavity. For the nRMSE metrics shown in Tables \ref{tab:ablation_study}, \ref{tab:sparse_data}, and \ref{tab:truncation}, we considered both velocity components and reported them as ($ \tilde{x} \pm \sigma_{\textnormal{rob}}$). Here, $ \tilde{x} $ signifies the median, while $ \sigma_{\text{rob}} = 1.4826 \times \textnormal{MAD} $ represents the robust standard deviation (std.), with MAD denoting the mean absolute deviation.

\section{Experimental Setup and Results}
\subsection{Dataset}
\subsubsection{Patient-specific Computational Fluid Dynamics (CFD) Heart Model}\label{sec:CFD_dataset}
To validate our approaches, we utilized a new patient-specific physiological CFD model of cardiac flow developed by the IMAG laboratory. This model features a more realistic mitral valve compared to the previous version \cite{chnafa_image-based_2014, chnafa_image-based_2016}. We followed the same method described in \cite[Sec. 2.3]{vixege_physics-constrained_2021} to generate 100 simulated Doppler images evenly distributed over a cardiac cycle, with a signal-to-noise ratio (SNR) equal to 50 dB. Each image comprised 80 scanlines with 200 samples per scanline. As Doppler power information was not available, we set the weight for the data fidelity term in both PINNs and \ivfm to one, \ie $ \omega = W = 1 $.

\begin{figure}
    \centering
    \includegraphics[width=0.47\textwidth]{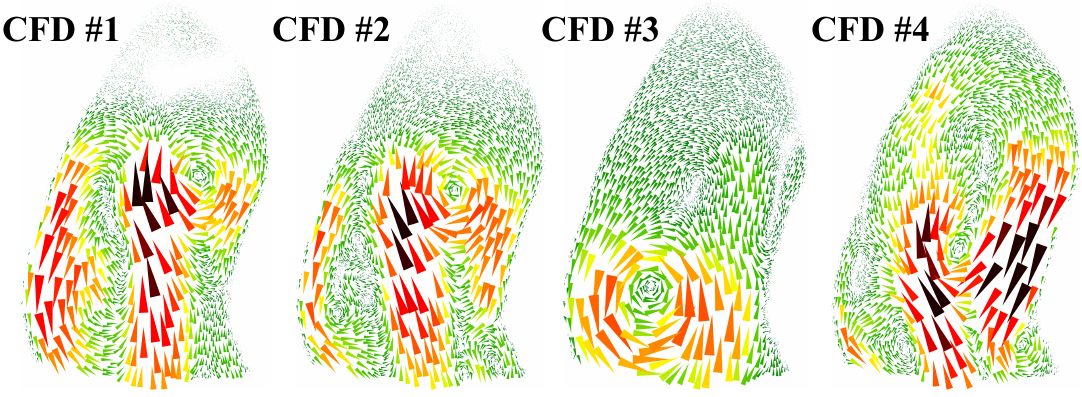}
    \caption{Simulated color Doppler image during early filling derived from patient-specific CFD heart models with four variants of mitral valves. CFD \#1-3 represent cases following mitral valve replacement with a bioprosthetic valve, while CFD \#4 is a normal case.}
    \label{fig:cfd}
\end{figure}

\subsubsection{CFD and \textit{In Vivo} Apical Three-Chamber (A3C) Training Dataset for \nnunet}\label{sec:train_dataset}
Based on the CFD model described in Section \ref{sec:CFD_dataset}, we introduced variations to the mitral valve geometry, by modifying opening angle and orientation, to simulate three distinct cases following mitral valve replacement with a bioprosthetic valve (CFD \#1-3) \cite{faludi_left_2010}. CFD \#1 and \#2 corresponded to two different inflow jet orientations, with a slight increase in mitral valve cross-sectional area in CFD \#2. CFD \#3 mimicked blood flow with a wide-opened mitral valve, resulting in a weak jet with limited penetration. CFD \#1-3 were included as training/validation data for the \nnunet, while the unaltered model representing a normal case (CFD \#4) was used for testing. Fig. \ref{fig:cfd} provides an example of a simulated color Doppler image during early filling for each CFD model. Although these four models shared the underlying cardiac geometry, modifications to the mitral valve resulted in sufficiently diverse intraventricular flows to reduce the training bias.

Due to the limited availability of simulated Doppler data for training a supervised model, we chose to include \invivo A3C duplex (B-mode + color Doppler) data in our training dataset. This decision was further elaborated in Section \ref{sec:pinnsVSnnunet}. These data aligned with the dataset of prior studies \cite{ling_dealiasing_2023, mehregan_doppler_2014}, acquired using a Vivid 7 ultrasound system (GE Healthcare, USA) with a GE 5S cardiac sector probe (bandwidth = 2--5 MHz). Further details about this dataset can be found in \cite[Sec. III-A.1]{ling_dealiasing_2023}.

We processed the \invivo A3C data to ensure high-quality training data. We initially filtered out low-quality data, resulting in a compilation of 92 Doppler echocardiographic cineloops from 37 patients, totaling 2668 frames. Subsequently, we performed preprocessing using \mbox{ASCENT} \cite{ling_extraction_2023, ling_dealiasing_2023}. This process involved segmenting the left ventricle cavity on B-mode images to define the region of interest and boundary conditions, which varied across frames, and correcting aliased pixels on the corresponding color Doppler images. Finally, we applied the \ivfm method to reconstruct the 2D vector field in the left ventricle, serving as the gold standard for training the \nnunet. For training purposes, the physics-constrained dataset was split subject-wise into 74/10/8 clinical cineloops plus 2/1/1 CFD simulations, resulting in 2037/434/197 \invivo and 200/100/100 simulated images for training/validation/testing.

\subsection{Training Strategies}\label{sec:training_strategies}
All methods were implemented in the same PyTorch-based framework to ensure consistent training and optimization. The training configurations for each approach were as follows:

\subsubsection{\rbpinns and \alpinns}\label{sec:training_strategies_pinns}
For both \rbpinns and \alpinns, we used a dual-stage optimization strategy, involving two stages with a total of $ I = 2500 $ iterations. In the first stage, we applied AdamW optimization for $ 0.9 \times I $ iterations, updating all learnable parameters with a learning rate of $ \eta_{\theta_{\textnormal{NN}}} = \eta_{\lambda} = \eta_{\mu} = 10^{-5} $. Then, in the fine-tuning stage, which comprised the remaining 10\% of the iterations, we utilized L-BFGS optimization. In this stage, only the network parameters were updated, while the learnable loss weights from the first stage were retained. The L-BFGS optimizer was configured with a maximum of ten iterations per optimization step, and the strong Wolfe line search conditions.

This strategy ensured a balance between accuracy and optimization duration, enhancing the stability and efficiency of PINNs' optimization process. The advantage of this strategy was further demonstrated in Section \ref{sec:pinns_ablation}.

\subsubsection{Physics-Guided \nnunet}
Our physics-guided \nnunet underwent 1000 epochs of training with the following configurations: a patch size of ($192 \times 40$) pixels, a batch size of 4, and SGD optimizer with an initial learning rate of 0.01, paired with a linear decay scheduler.

\begin{table}
    \centering
    \caption{Ablation study on 100 simulated Doppler images using \rbpinns. Optimization time is consistent (100 seconds per frame) across all configuration combinations. Default settings are highlighted in \colorbox{blue!15}{purple}.}
    \begin{NiceTabular}{@{}ccccc@{}}
        \CodeBefore
            \cellcolor{blue!15}{6-3, 6-4, 6-5}
        \Body
        \toprule
        \Block[C]{2-1}{Pre-optimized \\ weights}
        &
        \Block[C]{2-1}{Dual-stage \\ optimization}
        & 
        \Block[C]{1-2}{$ r^2 (\uparrow) $}
        & &
        \Block[C]{2-1}{nRMSE $ [\%] (\downarrow) $ \\ ($ \tilde{x} \pm \sigma_\textnormal{rob}$)}\\
        \cmidrule(lr){3-4}
         & & $ v_r $ & $ v_{\theta} $ & \\
        \midrule
        \xmark & \xmark & 0.88 & 0.23 & $ 4.3 \pm 2.2 $\\
        \xmark & \cmark & 0.97 & 0.57 & $ 2.8 \pm 1.2 $\\
        \cmark & \xmark & 0.96 & 0.58 & $ 2.4 \pm 1.0 $\\
        \cmark & \cmark & \textbf{0.99} & \textbf{0.66} & $ \bm{2.2} \pm 1.0 $\\
        \bottomrule
    \end{NiceTabular}
    \label{tab:ablation_study}
\end{table}

\begin{figure}
    \centering
    \includegraphics[width=0.48\textwidth]{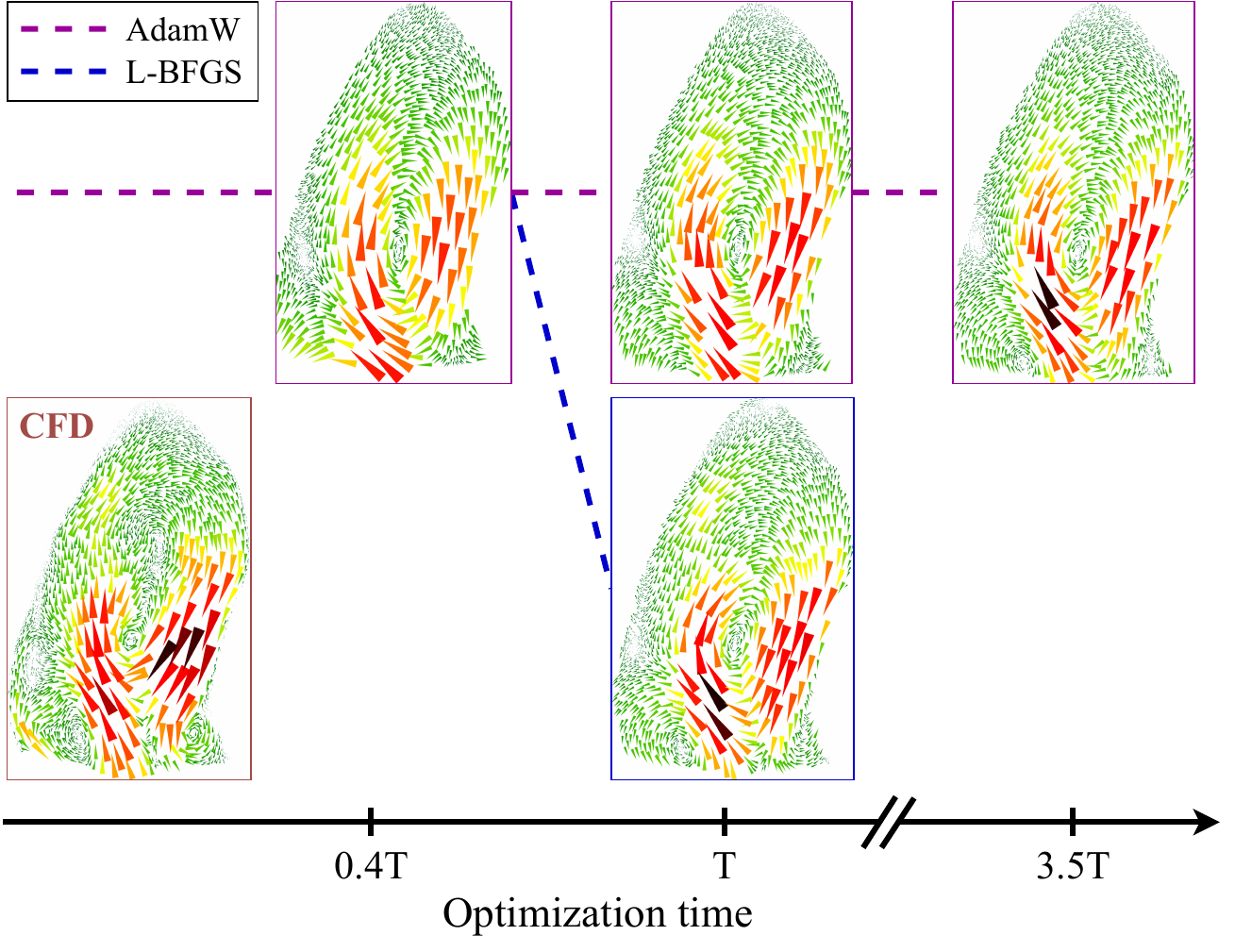}
    \caption{Dual-stage (AdamW + L-BFGS) versus single-stage (AdamW only) optimization using \rbpinns initialized with pre-optimized weights. T refers to the total amount of time required for dual-stage optimization. In this example, 3.5 $ \times $ more time is needed for single-stage optimization (top right) to converge to a similar solution given by dual-stage optimization (bottom right).}
    \label{fig:dual_stage}
\end{figure}

\begin{figure*}
    \centering
    \includegraphics[width=0.92\textwidth]{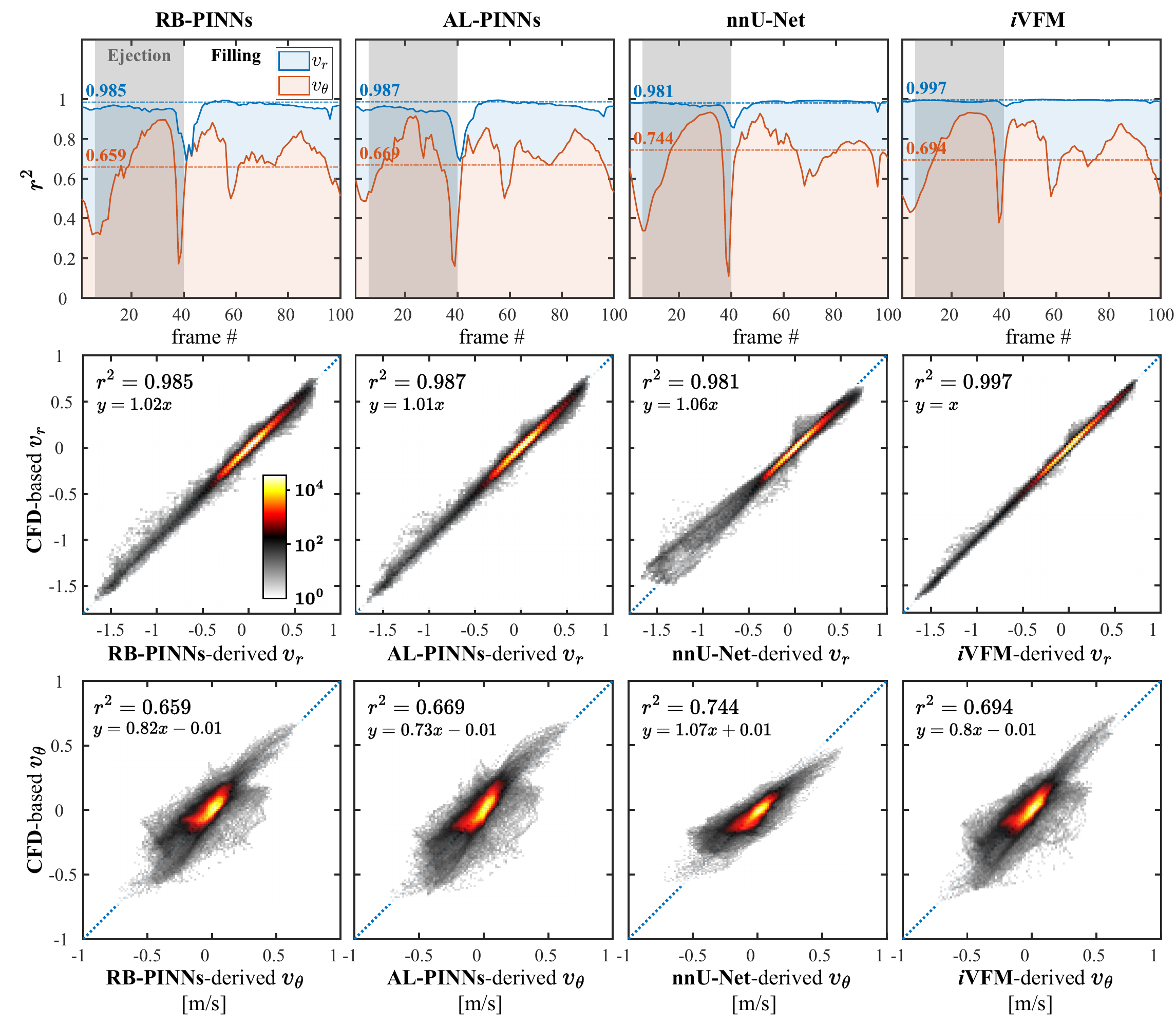}
    \caption{Top row: time-varying squared correlation between CFD-based velocities and reconstructed velocities by each method; mid and bottom rows: CFD-based velocities versus estimated velocities derived from various methods. For mid and bottom rows, velocity data from 100 simulated color Doppler images were pooled. The binned scatter plots show the number of velocity occurrences.}
    \label{fig:r2}
\end{figure*}

\subsection{Experimental Results}
\subsubsection{Pre-Optimized Weights and Dual-Stage Optimization Enhanced PINNs' Performance}\label{sec:pinns_ablation}
The ablation study presented in Table \ref{tab:ablation_study} highlights that the combination of pre-optimized weights and dual-stage optimization in \rbpinns yielded the best performance within a fixed optimization time. The dual-stage optimization strategy significantly reduced the optimization time of PINNs methods. Figure \ref{fig:dual_stage} provides a visual representation of a case where optimization was conducted using pre-optimized weights with and without dual-stage optimization. In this example, single-stage optimization with AdamW required $3.5 \times$ more optimization time to achieve a visually similar solution compared to dual-stage optimization. Subsequent experiments with PINNs followed this optimization strategy—dual-stage optimization with pre-optimized weight initialization.

\subsubsection{NN-based Approaches Aligned with the Original \ivfm}
\begin{figure*}
    \centering
    \includegraphics[width=0.95\textwidth]{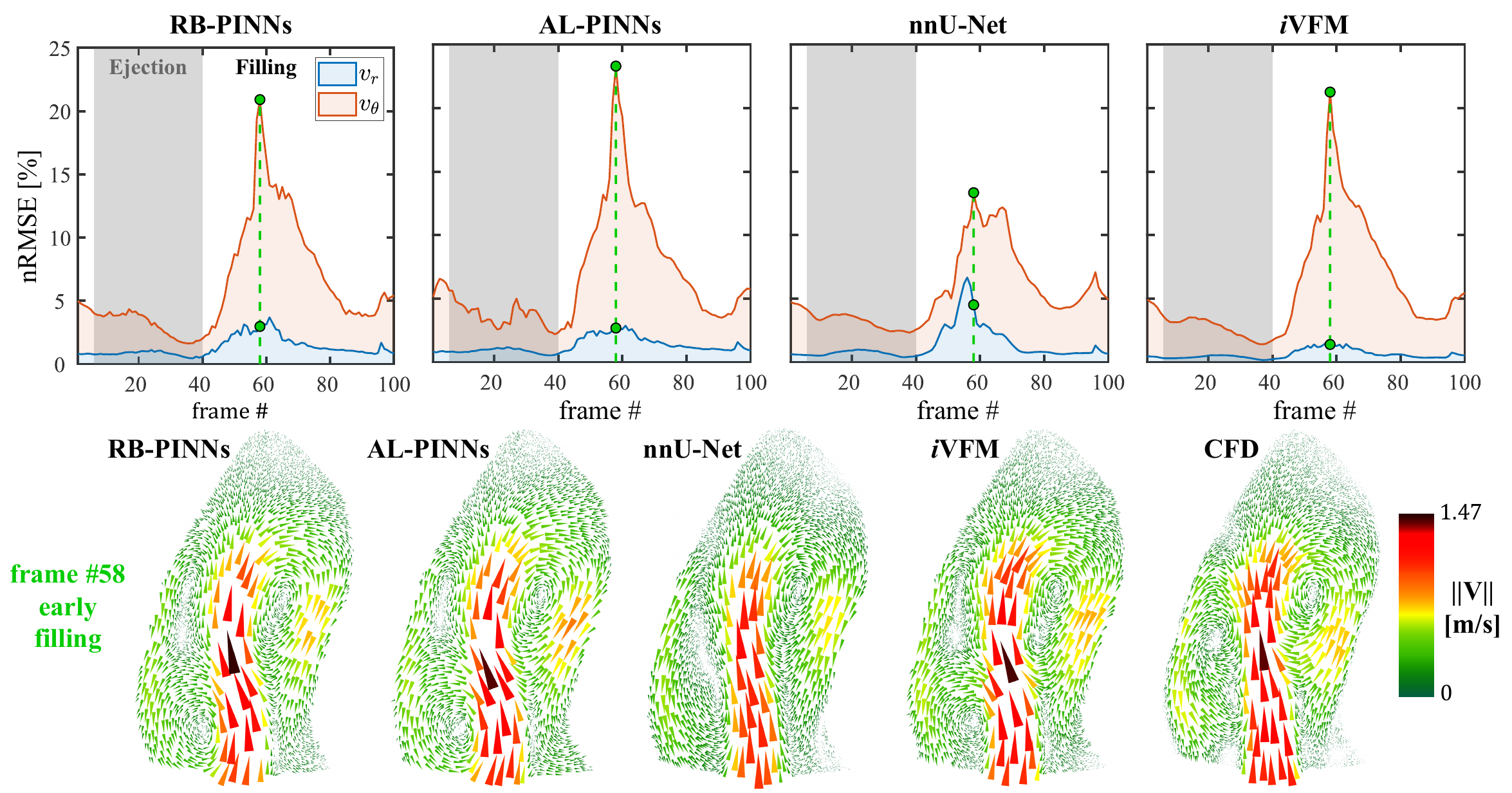}
    \caption{Normalized root-mean-square errors (nRMSE) between CFD-based and estimated velocity vectors by different techniques.}
    \label{fig:nRMSE}
\end{figure*}

All methods achieved high correlation in the radial velocity estimation on the 100 simulated color Doppler images derived from CFD \#4, with $ r^2_{v_r} > 0.98 $ (see Fig. \ref{fig:r2}). For angular velocity correlation, both PINNs, \rbpinns and \alpinns performed similarly to \ivfm, $ r^2_{v_{\theta}} = 0.659 $ and $ 0.669 $ versus $ 0.694 $, while \nnunet surpassed \ivfm ($ r^2_{v_{\theta}} = 0.744 $). This suggests the effectiveness of a supervised approach in learning intraventricular blood flow patterns. However, \nnunet tended to be less precise when estimating highly negative radial velocities on simulated Doppler data, potentially due to the limited CFD training samples. Interestingly, all NN methods exhibited more errors for the radial component than \ivfm, highlighting the high robustness and precision of the physics-constrained \ivfm approach (see Fig. \ref{fig:nRMSE}). The nRMSE of \ivfm ranged between 0.2\%-1.6\% and 1.4\%-21.3\% for the radial and angular velocities, respectively. Among all methods, \alpinns had the highest nRMSE for angular velocities (2.2\% to 23.3\%). Despite having the highest nRMSE for radial velocities (3.8\% to 6.7\%), \nnunet produced the least errors in angular velocity estimation (2.3\% to 13.3\%). A cineloop showing the reconstructed field by PINNs and \nnunet versus CFD can be found in the Supplementary Material \href{https://www.youtube.com/watch?v=cdjIo9sP6pA}{\color{Red} \faYoutube}.

The final optimized values of the penalty coefficients were ($ \textnormal{median}  \pm  \textnormal{robust std.} $ [min, max]): $ \mu_1 = 0.90 \pm  0.11 \: [0.32, 1.00] $, $ \mu_2 = 0.90 \pm  0.12 \: [0.32, 1.02] $, $ \mu_3 = 1.19 \pm  0.22 \: [0.99, 2.36] $ for \rbpinns, and $ \mu = 2.02 \pm  0.01 \: [2.01, 2.07] $ for \alpinns.

\begin{table*}
    \parbox[t]{.55\linewidth}{
    \centering
    \caption{Metrics computed on 100 full and sparse simulated Doppler images}
    \begin{threeparttable}
    \begin{tabular}{@{}l ccc ccc@{}}
        \toprule
        \multirow{3}[3]{*}{Methods}
        &
        \multicolumn{3}{c}{Full data}
        &
        \multicolumn{3}{c}{Sparse data$^*$}\\
        \cmidrule(lr){2-4} \cmidrule(lr){5-7}
         & \multicolumn{2}{c}{$ r^2 (\uparrow) $} & \multirow{2}[2]{*}{\begin{tabular}[c]{@{}c@{}}nRMSE $ [\%] (\downarrow) $ \\
        ($ \tilde{x} \pm \sigma_\textnormal{rob}$)\end{tabular}} & \multicolumn{2}{c}{$ r^2 (\uparrow) $} & \multirow{2}[2]{*}{\begin{tabular}[c]{@{}c@{}}nRMSE $ [\%] (\downarrow) $ \\
        ($ \tilde{x} \pm \sigma_\textnormal{rob}$)\end{tabular}} \\
        \cmidrule(lr){2-3} \cmidrule(lr){5-6}
         & $ v_r $ & $ v_{\theta} $ & & $ v_r $ & $ v_{\theta} $ & \\
        \midrule
        \ivfm               & \textbf{1.00} & 0.69 & $ \bm{1.7} \pm 1.0 $  & 0.85 & 0.58 & $ 3.2 \pm 2.3 $\\
        
        \midrule
        
        \rbpinns            & 0.99 & 0.66 & $ 2.2 \pm 1.0 $  & 0.86 & 0.56 & $ 3.4 \pm 1.6 $\\
        \alpinns            & 0.99 & 0.67 & $ 2.5 \pm 1.0 $ & 0.80 & 0.58 & $ 3.5 \pm 1.4 $\\
        
        \midrule
      
        $\nnunet^{\ddagger}$ & 0.99 & 0.60 & $ 2.3 \pm 0.9 $  & 0.67 & 0.12 & $ 6.0 \pm 3.0 $\\
        $\nnunet^{\dagger}$ & 0.99 & \textbf{0.74} & $ 2.1 \pm 0.9 $  & 0.64 & 0.49 & $ 6.0 \pm 3.3 $\\
        \nnunet             & 0.98 & \textbf{0.74} & $ 2.1 \pm 0.9 $  & \textbf{0.88} & \textbf{0.71} & $ \bm{2.4} \pm 1.0 $\\
        \bottomrule
    \end{tabular}
    \begin{tablenotes}[para,flushleft]
        * indicates data masked every 9 out of 10 scanlines from the center to the borders.\\
        $\ddagger$ means training without both physical regularization term (PDE loss) and random scanline masking augmentation.\\
        $\dagger$ signifies training with physical regularization term (PDE loss), but without random scanline masking augmentation.
    \end{tablenotes}
    \end{threeparttable}
    \label{tab:sparse_data}
    }
    \hfill
    \parbox[t]{.45\linewidth}{
    \centering
    \caption{Comparison of training, optimization, and inference times for NN-based methods and \ivfm}
    \begin{NiceTabular}{@{}lcccc@{}}
        \toprule
        Methods
        &
        Device
        &
        \Block[C]{}{No. trainable \\ parameters}
        & 
        \Block[C]{}{Training \\ time}
        & 
        \Block[C]{}{Optimization/ \\ inference time \\ per frame}
        \\
        \midrule
        \ivfm    & CPU & -      & -    & 0.2 s \\
        \midrule
        \rbpinns & GPU & 18.6 k & -    & 100 s  \\
        \alpinns & GPU & 18.6 k & -    & 100 s  \\
        \midrule
        \nnunet  & GPU & 7 M    & 12 h & \textbf{0.05 s} \\
        \bottomrule
    \end{NiceTabular}
    \label{tab:inference_time}
    }
\end{table*}

\subsubsection{\nnunet Demonstrated Better Generalizability and Robustness on Sparse Doppler Data}
Table \ref{tab:sparse_data} presents metrics for each method on both full and sparse simulated Doppler images. In the evaluation on full data, \ivfm achieved the highest correlation for radial velocities and the lowest nRMSE, while \nnunet, trained with the physical regularization term (PDE loss), excelled in the correlation of angular velocities. This implies that incorporating physical regularization helps constrain the \nnunet's output to better adhere to the laws of physics.

As expected, the performance of all methods was significantly impacted when evaluated on sparse data, where nine out of ten scanlines were masked. Remarkably, \nnunet, which was trained with both the physical regularization term and the random scanline masking augmentation, demonstrated the least decline in performance. It maintained a high correlation for both radial and angular velocities while achieving the least nRMSE. This finding underscores the benefit of this augmentation in supervised learning for enhanced generalization. Despite having lower correlations than \nnunet, \ivfm remained robust, producing a lower nRMSE than \rbpinns and \alpinns.

\subsubsection{\nnunet Exhibited Superior Reconstruction Speed}
Table \ref{tab:inference_time} provides a comparison of the training, optimization, and inference times for NN-based approaches against \ivfm. These metrics were computed using a 16 GB V100 GPU for NN methods and an Intel i5-11500H CPU for \ivfm. As the only supervised approach in the comparison, \nnunet required 12 hours of training but achieved the fastest per-frame inference time, taking only 0.05 second. \ivfm ranked second, with a reconstruction time of 0.2 second per frame. Notably, both PINNs necessitated longer optimization times, around 100 seconds, which is a recognized drawback of this approach.

\begin{figure*}[t]
    \centering
    \includegraphics[width=0.92\textwidth]{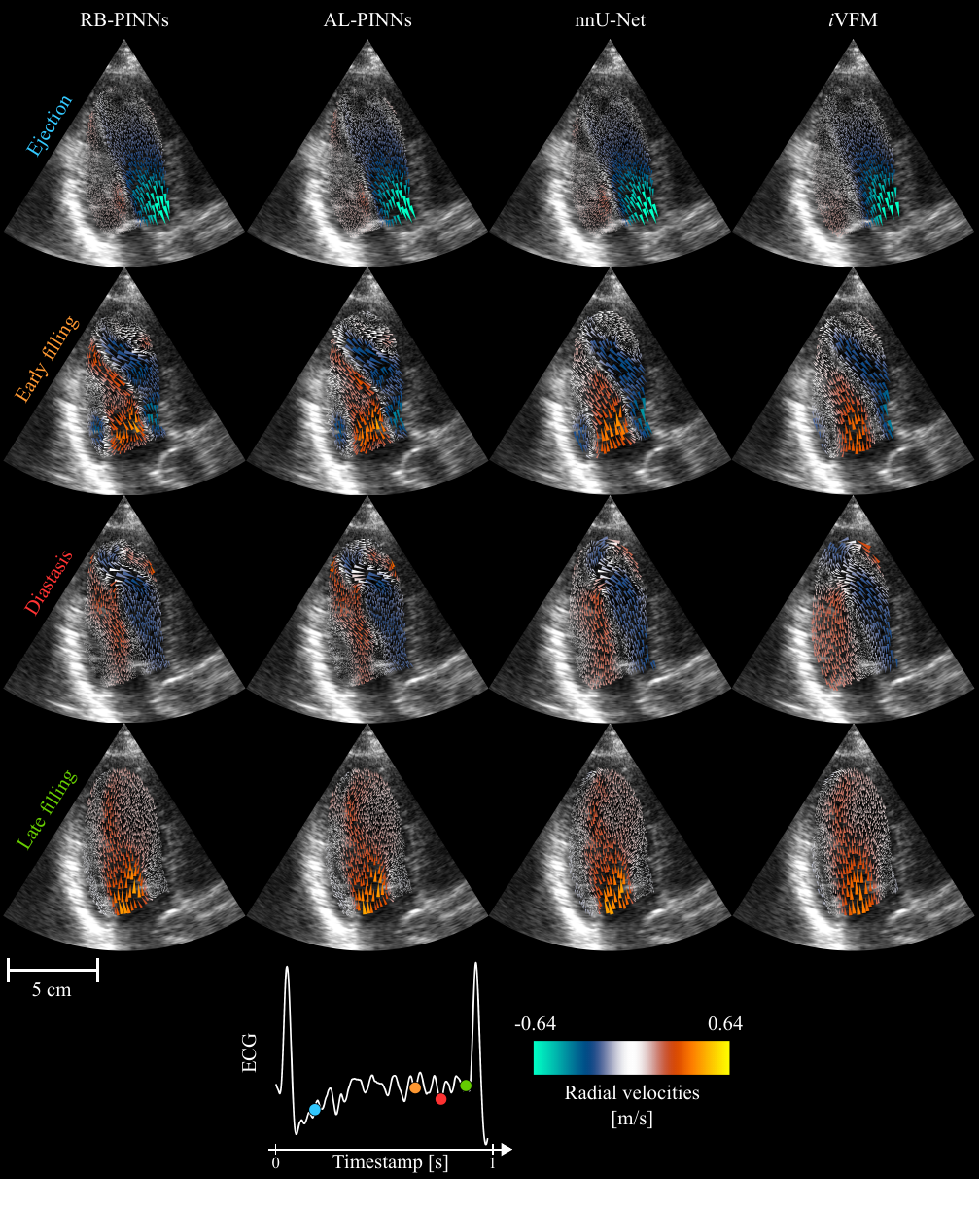}
    \caption{Reconstruction of intraventricular vector blood flow in a patient using NN-based approaches and \ivfm. The color of the arrows represents the estimated radial velocity fields.}
    \label{fig:in_vivo}
\end{figure*}

\begin{figure*}
    \centering
    \includegraphics[width=0.75\textwidth]{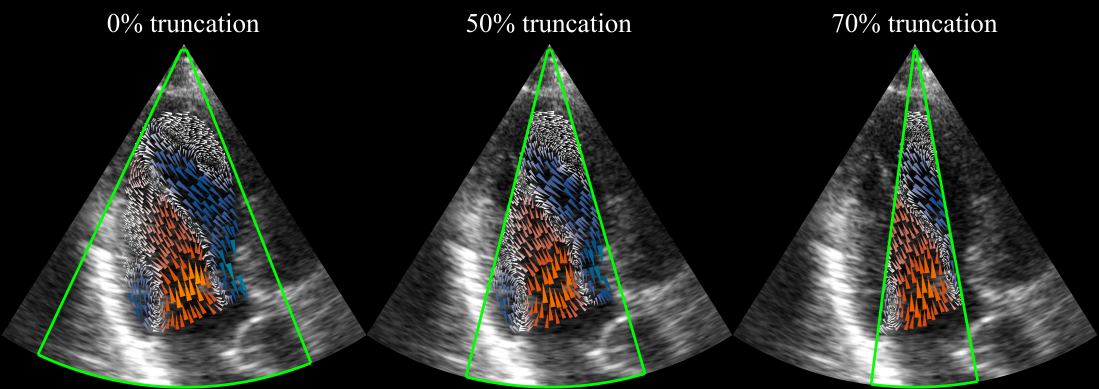}
    \caption{Intraventricular vector blood flow reconstruction from Doppler data with varying percentages of scanline truncation using physics-guided \nnunet. The color of the arrows represents the estimated radial velocity fields.}
    \label{fig:truncate}
\end{figure*}

\subsubsection{Clinical Application of Vector Blood Flow Mapping}
Fig. \ref{fig:in_vivo} showcases the intraventricular vector blood flow mapping by various methods on an \textit{in vivo} case at different cardiac phases, including ejection, early filling, diastasis, and late filling. The reconstructed flow patterns by all methods appear relatively similar, with \ivfm generating the smoothest flow patterns. Although the prominent vortex was less visible during early filling, it became more pronounced at the center of the left ventricle cavity during diastasis. Another example of vector blood flow reconstruction by all four methods on \invivo data is given in the Supplementary Material \href{https://www.youtube.com/watch?v=7K9z92SIy1M}{\color{Red} \faYoutube}.

\subsubsection{Robustness of \nnunet on Truncated Clinical Doppler Data}
Unlike other methods that required explicit boundary conditions, \nnunet learned these conditions implicitly during training. This offered an advantage as it reduced the need for specific knowledge about the flow at the endocardium. Table \ref{tab:truncation} illustrates \nnunet's behavior under Doppler scanline truncation, achieved by progressively cutting scanlines from both sides towards the center. The metrics were computed within the common region remaining after truncation. The results show stable performance up to a 50\% reduction. However, beyond this threshold, a more pronounced decrease in performance was observed. Fig. \ref{fig:truncate} visually demonstrates \nnunet's ability to consistently produce accurate intraventricular vector blood flow reconstructions, even with a significant 70\% truncation.

\section{Discussions}
Our study introduces alternative approaches to the physics-constrained \ivfm algorithm \cite{vixege_physics-constrained_2021}, leveraging the power of NNs: PINNs (\rbpinns and \alpinns) and a physics-guided supervised technique (\nnunet). These methods offer distinct strategies for the inherent constrained optimization problem in \ivfm.

In the PINNs framework, we addressed the same optimization problem as in \ivfm, but solved it differently using gradient descent and NNs. By incorporating governing equations, such as mass conservation and boundary conditions, PINNs inherently enforce physical laws during optimization, potentially leading to optimized intraventricular vector velocity fields.

On the other hand, the supervised approach (\nnunet) was trained on patient-specific CFD-derived simulations and \ivfm-estimated velocity field on \invivo Doppler data. The network learned the underlying flow patterns while adhering to physical principles through the use of physics-constrained labels and the physical regularization term in the loss function. This approach demonstrated robustness to data limitations, such as missing scanlines.

\begin{table}
    \centering
    \caption{\nnunet's metrics computed on 8 \invivo test cineloops of 197 frames with different percentages of scanline truncation}
    \begin{threeparttable}
        \begin{tabular}{@{}cccc@{}}
            \toprule
            \multirow{2}[2]{*}{
            \begin{tabular}[c]{@{}c@{}}Percentage of \\ 
            truncation [\%]\end{tabular}}
            &
            \multicolumn{2}{c}{$ r^2 (\uparrow) $}
            &
            \multirow{2}[2]{*}{
            \begin{tabular}[c]{@{}c@{}}nRMSE $ [\%] (\downarrow) $ \\
            ($ \tilde{x} \pm \sigma_\textnormal{rob}$)\end{tabular}
            }\\
            \cmidrule(lr){2-3}
             & $ v_r $ & $ v_{\theta} $ & \\
            \midrule
            20 & 0.97 & 0.96 & $ 2.4 \pm 1.2 $\\
            40 & 0.99 & 0.92 & $ 3.7 \pm 1.5 $\\
            50 & 0.99 & 0.87 & $ 4.3 \pm 1.6 $\\
            60 & 0.97 & 0.75 & $ 6.5 \pm 2.7 $\\
            70 & 0.94 & 0.59 & $ 8.4 \pm 2.9 $\\
            \bottomrule
        \end{tabular}
        \begin{tablenotes}[para,flushleft]
        Note: Comparison made with \nnunet's estimated velocity fields on full scanline data.
    \end{tablenotes}
    \end{threeparttable}
    \label{tab:truncation}
\end{table}

\subsection{PINNs versus Physics-Guided \nnunet} \label{sec:pinnsVSnnunet}
The application of PINNs deviates from the conventional optimization methods by leveraging NNs to find the optimal solution, which can be advantageous in complex physical problems. Although PINNs may not necessarily outperform analytical or numerical methods in computational efficiency, approximation accuracy, or convergence guarantees \cite{grossmann_can_2023}, they offer a unique advantage in terms of flexibility. This flexibility allows their architecture to remain relatively consistent across various physical optimization problems by adapting the loss functions to be optimized.

In our case of intraventricular blood flow reconstruction, we successfully improved the computational efficiency of PINNs while maintaining accuracy comparable to \ivfm. This was achieved by implementing a dual-stage optimization with the use of pre-optimized weights. For future exploration, both PINNs architectures could benefit from imposing hard boundary conditions rather than optimizing them in the form of soft constraints. Additionally, better strategies for automatically determining or learning the optimal smoothing regularization weight, rather than relying on heuristic search methods, will be investigated to further improve the robustness of PINNs.

Unlike PINNs, \nnunet operates within a supervised learning framework, heavily relying on labeled training data. In this study, CFD-derived simulations played a crucial role in providing ground truth velocity fields for training. This explains the high squared correlation achieved by \nnunet, as the simulated training samples shared the exact heart geometry despite variations in flow patterns due to different mitral valve conditions.

However, when trained exclusively on simulated data, \nnunet struggled to correctly estimate vector blood flow in \textit{in vivo} color Doppler data due to a distribution shift between the simulated and real data. This arose from limitations in the current physiological spectrum of the simulations. To bridge this gap, we included \ivfm-estimated velocities in our training data, allowing  \nnunet to learn from solutions representative of clinical Doppler data generated by the established \ivfm method. As shown in Fig. \ref{fig:in_vivo}, \nnunet effectively learned the underlying flow pattern and physical properties, \ie the free-divergence and boundary conditions, from the training samples generated by \ivfm. 

Moreover, as illustrated in Fig. \ref{fig:truncate}, \nnunet can precisely reconstruct intraventricular flow on truncated Doppler data, where PINNs and \ivfm cannot be directly applied as such due to incomplete and unknown boundary conditions. This potentially makes \nnunet the preferred candidate for clinical applications, especially considering that most clinical Doppler acquisitions do not capture the entire left ventricular cavity due to limitations in probe placement or patient anatomy. With the added advantage of the shortest inference time, \nnunet has real-time capabilities suitable for clinical settings. Future work will focus on generating more patient-specific CFD models and creating more realistic simulated Doppler data \cite{sun_pipeline_2022} to avoid the bias associated with using \ivfm estimates as a reference to train our model.

\subsection{Limitations of Color Doppler and Vector Flow Mapping}
Conventional color Doppler echocardiography is subject to various limitations, posing challenges for accurate vector flow mapping. These limitations include clutter signals arising from myocardial tissue and valve leaflets, aliasing artifacts caused by Doppler velocity overshooting beyond Nyquist velocity, and low spatial and temporal resolutions. While the impact of clutter signal filtering on flow reconstruction is acknowledged, its specific effects were not investigated in this study. Our input data were already clutter-filtered (\textit{in vivo} non scan-converted data from a GE scanner) or clutter-free (simulated data). Some researchers address clutter filtering in color flow imaging using deep learning (DL) techniques \cite{wang_preliminary_2021}.

The clinical color Doppler data used for training and testing in this study contained only single aliasing, corrected by a DL-based unwrapping algorithm from our previous work \cite{ling_dealiasing_2023}. It is important to note that this algorithm may face limitations in scenarios involving multiple aliasing, particularly in valvular disease. A potential solution could involve training a supervised DL model with multi-aliased data and their alias-free labels, such as in interferometric imaging \cite{spoorthi_phasenet2.0_2020} or color Doppler imaging of the femoral bifurcation \cite{nahas_deep_2020}.

The temporal resolution of clinical color Doppler, typically ranging from 10 to 15 frames per second, significantly restricts the use of temporal information. Consequently, the application of physical constraints is limited to those that are not time-dependent, such as mass conservation for an incompressible fluid. Around 25 harmonics are necessary to accurately record pressure time derivatives within the left ventricle with a 5\% margin of error \cite{krovetz_frequency_1974}. Hence, a color Doppler frame rate of 25 should be ideally sought to characterize intracardiac blood flow. Although we did not address such a strategy in this study, one approach would be to use Doppler information from two or three successive cardiac cycles. To overcome these limitations, advances in increasing the frame rate of color Doppler imaging, such as diverging scan sequences \cite{posada_staggered_2016, ramalli_high-frame-rate_2020}, have been explored. Another promising avenue involves utilizing multi-line transmission \cite{ramalli_real-time_2018}, especially given the robust performance of our models on sparse Doppler data, notably \nnunet. Higher frame rates offer the possibility of incorporating more complex physical constraints, including vorticity, Euler, or Navier-Stokes equations, potentially enhancing flow reconstruction accuracy. While this might pose challenges for the original \ivfm due to the non-linear terms in the equations, it aligns seamlessly with PINNs, requiring minimal changes to the loss function.

\subsection{Future Directions}
The successful application of PINNs in mapping intraventricular vector flow from color Doppler paves the way for future investigations. Upcoming research will prioritize the integration of high-frame-rate color Doppler with PINNs while incorporating the governing Navier-Stokes equations. This combination aims to leverage the temporal information to obtain a more accurate velocity field and the pressure gradient within the left ventricle.

Furthermore, with the development of our fully automated and robust tools, including left ventricular segmentation, dealiasing, and velocity field reconstruction using NNs, we anticipate extracting potential biomarkers from intracardiac vector blood flow for enhanced clinical insights and diagnostic capabilities.

\section{Conclusion}
Our study presents novel approaches based on NNs for intraventricular vector flow mapping, utilizing gradient-based optimization through PINNs (\rbpinns and \alpinns) or physics-guided supervised learning (\nnunet). These methods offer contrasting strategies to tackle the ill-posed inverse problem for vector flow reconstruction. Our results demonstrate the remarkable capabilities of both PINNs and \nnunet in reconstructing intraventricular vector blood flow fields. Particularly noteworthy is \nnunet's performance, demonstrating quasi-real-time capability, robustness on sparse Doppler data, and independence from explicit boundary conditions. These characteristics position \nnunet as a promising solution for real-time clinical applications.

\section*{Acknowledgment}
Olivier Bernard and Damien Garcia share the last authorship for this work. The authors thank François Tournoux (Montreal University—CHUM—CRCHUM) for providing access to the \invivo data used in this study.

\bibliographystyle{IEEEtran}
\bibliography{tuffc.bib}

\begin{thebibliography}{10}
\providecommand{\url}[1]{#1}
\csname url@samestyle\endcsname
\providecommand{\newblock}{\relax}
\providecommand{\bibinfo}[2]{#2}
\providecommand{\BIBentrySTDinterwordspacing}{\spaceskip=0pt\relax}
\providecommand{\BIBentryALTinterwordstretchfactor}{4}
\providecommand{\BIBentryALTinterwordspacing}{\spaceskip=\fontdimen2\font plus
\BIBentryALTinterwordstretchfactor\fontdimen3\font minus \fontdimen4\font\relax}
\providecommand{\BIBforeignlanguage}[2]{{%
\expandafter\ifx\csname l@#1\endcsname\relax
\typeout{** WARNING: IEEEtran.bst: No hyphenation pattern has been}%
\typeout{** loaded for the language `#1'. Using the pattern for}%
\typeout{** the default language instead.}%
\else
\language=\csname l@#1\endcsname
\fi
#2}}
\providecommand{\BIBdecl}{\relax}
\BIBdecl

\bibitem{vixege_physics-constrained_2021}
F.~Vixège \emph{et~al.}, ``\BIBforeignlanguage{en}{Physics-constrained intraventricular vector flow mapping by color {Doppler}},'' \emph{\BIBforeignlanguage{en}{Phys. Med. Biol.}}, vol.~66, no.~24, p. 245019, Dec. 2021.

\bibitem{assi_intraventricular_2017}
K.~C. Assi \emph{et~al.}, ``\BIBforeignlanguage{eng}{Intraventricular vector flow mapping-a {Doppler}-based regularized problem with automatic model selection},'' \emph{\BIBforeignlanguage{eng}{Phys. Med. Biol.}}, vol.~62, no.~17, pp. 7131--7147, Aug. 2017.

\bibitem{asami_accuracy_2017}
R.~Asami, T.~Tanaka, K.-i. Kawabata, K.~Hashiba, T.~Okada, and T.~Nishiyama, ``\BIBforeignlanguage{en}{Accuracy and limitations of vector flow mapping: left ventricular phantom validation using stereo particle image velocimetory},'' \emph{\BIBforeignlanguage{en}{J. Echocardiogr.}}, vol.~15, no.~2, pp. 57--66, Jun. 2017.

\bibitem{meyers_colour-doppler_2020}
B.~A. Meyers, C.~J. Goergen, P.~Segers, and P.~P. Vlachos, ``Colour-{Doppler} echocardiography flow field velocity reconstruction using a streamfunction–vorticity formulation,'' \emph{J. R. Soc. Interface}, vol.~17, no. 173, p. 20200741, Dec. 2020.

\bibitem{daae_intraventricular_2021}
A.~S. Daae, M.~S. Wigen, S.~Fadnes, L.~Løvstakken, and A.~Støylen, ``Intraventricular {Vector} {Flow} {Imaging} with {Blood} {Speckle} {Tracking} in {Adults}: {Feasibility}, {Normal} {Physiology} and {Mechanisms} in {Healthy} {Volunteers},'' \emph{Ultrasound Med. Biol.}, vol.~47, no.~12, pp. 3501--3513, Dec. 2021.

\bibitem{raissi_physics-informed_2019}
M.~Raissi, P.~Perdikaris, and G.~E. Karniadakis, ``Physics-informed neural networks: {A} deep learning framework for solving forward and inverse problems involving nonlinear partial differential equations,'' \emph{J. Comput. Phys.}, vol. 378, pp. 686--707, Feb. 2019.

\bibitem{baydin_automatic_2018}
A.~Baydin, B.~Pearlmutter, A.~Radul, and J.~Siskind, ``Automatic differentiation in machine learning: {A} survey,'' \emph{J. Mach. Learn. Res.}, vol.~18, pp. 1--43, Apr. 2018.

\bibitem{cai_physics-informed_2021}
S.~Cai, Z.~Mao, Z.~Wang, M.~Yin, and G.~E. Karniadakis, ``\BIBforeignlanguage{en}{Physics-informed neural networks ({PINNs}) for fluid mechanics: a review},'' \emph{\BIBforeignlanguage{en}{Acta Mech. Sin.}}, vol.~37, no.~12, pp. 1727--1738, Dec. 2021.

\bibitem{arzani_uncovering_2021}
A.~Arzani, J.-X. Wang, and R.~M. D'Souza, ``Uncovering near-wall blood flow from sparse data with physics-informed neural networks,'' \emph{Phys. Fluids}, vol.~33, no.~7, p. 071905, Jul. 2021.

\bibitem{kissas_machine_2020}
G.~Kissas, Y.~Yang, E.~Hwuang, W.~R. Witschey, J.~A. Detre, and P.~Perdikaris, ``Machine learning in cardiovascular flows modeling: {Predicting} arterial blood pressure from non-invasive {4D} flow {MRI} data using physics-informed neural networks,'' \emph{Comput. Methods Appl. Mech. Eng.}, vol. 358, p. 112623, Jan. 2020.

\bibitem{wang_physics-informed_2023}
L.~Wang, H.~Wang, L.~Liang, J.~Li, Z.~Zeng, and Y.~Liu, ``Physics-informed neural networks for transcranial ultrasound wave propagation,'' \emph{Ultrasonics}, vol. 132, p. 107026, Jul. 2023.

\bibitem{yin_swenet_2023}
Z.~Yin, G.-Y. Li, Z.~Zhang, Y.~Zheng, and Y.~Cao, ``{SWENet}: a physics-informed deep neural network ({PINN}) for shear wave elastography,'' \emph{IEEE Trans. Med. Imaging}, pp. 1--1, 2023.

\bibitem{guan_towards_2023}
H.~Guan, J.~Dong, and W.-N. Lee, ``Towards {Real}-time {Training} of {Physics}-informed {Neural} {Networks}: {Applications} in {Ultrafast} {Ultrasound} {Blood} {Flow} {Imaging},'' Sep. 2023, arXiv:2309.04755 [cs.CE].

\bibitem{faroughi_physics-guided_2024}
S.~A. Faroughi \emph{et~al.}, ``Physics-{Guided}, {Physics}-{Informed}, and {Physics}-{Encoded} {Neural} {Networks} and {Operators} in {Scientific} {Computing}: {Fluid} and {Solid} {Mechanics},'' \emph{J. Comput. Inf. Sci. Eng.}, vol.~24, no. 040802, Jan. 2024.

\bibitem{isensee_nnu-net_2021}
F.~Isensee, P.~F. Jaeger, S.~A.~A. Kohl, J.~Petersen, and K.~H. Maier-Hein, ``\BIBforeignlanguage{en}{{nnU}-{Net}: a self-configuring method for deep learning-based biomedical image segmentation},'' \emph{\BIBforeignlanguage{en}{Nat. Methods}}, vol.~18, no.~2, pp. 203--211, 2021.

\bibitem{chen_gradnorm_2018}
Z.~Chen, V.~Badrinarayanan, C.-Y. Lee, and A.~Rabinovich, ``{GradNorm}: {Gradient} {Normalization} for {Adaptive} {Loss} {Balancing} in {Deep} {Multitask} {Networks},'' in \emph{Proc. ICML}, Jul. 2018, pp. 794--803.

\bibitem{heydari_softadapt_2019}
A.~A. Heydari, C.~A. Thompson, and A.~Mehmood, ``{SoftAdapt}: {Techniques} for {Adaptive} {Loss} {Weighting} of {Neural} {Networks} with {Multi}-{Part} {Loss} {Functions},'' Dec. 2019, arXiv:1912.12355 [cs.LG].

\bibitem{bischof_multi-objective_2021}
R.~Bischof and M.~Kraus, ``Multi-{Objective} {Loss} {Balancing} for {Physics}-{Informed} {Deep} {Learning},'' Nov. 2021, arXiv:2110.09813 [cs.LG].

\bibitem{lu_physics-informed_2021}
L.~Lu, R.~Pestourie, W.~Yao, Z.~Wang, F.~Verdugo, and S.~G. Johnson, ``Physics-{Informed} {Neural} {Networks} with {Hard} {Constraints} for {Inverse} {Design},'' \emph{SIAM J. Sci. Comput.}, vol.~43, no.~6, pp. B1105--B1132, Jan. 2021.

\bibitem{son_enhanced_2023}
H.~Son, S.~W. Cho, and H.~J. Hwang, ``Enhanced physics-informed neural networks with {Augmented} {Lagrangian} relaxation method ({AL}-{PINNs}),'' \emph{Neurocomputing}, vol. 548, p. 126424, Sep. 2023.

\bibitem{berg_unified_2018}
J.~Berg and K.~Nyström, ``A unified deep artificial neural network approach to partial differential equations in complex geometries,'' \emph{Neurocomputing}, vol. 317, pp. 28--41, Nov. 2018.

\bibitem{sukumar_exact_2022}
N.~Sukumar and A.~Srivastava, ``Exact imposition of boundary conditions with distance functions in physics-informed deep neural networks,'' \emph{Comput. Methods Appl. Mech. Eng.}, vol. 389, p. 114333, Feb. 2022.

\bibitem{hestenes_multiplier_1969}
M.~R. Hestenes, ``\BIBforeignlanguage{en}{Multiplier and gradient methods},'' \emph{\BIBforeignlanguage{en}{J. Optim. Theory Appl.}}, vol.~4, no.~5, pp. 303--320, Nov. 1969.

\bibitem{loshchilov_decoupled_2018}
I.~Loshchilov and F.~Hutter, ``Decoupled weight decay regularization,'' in \emph{Proc. ICLR}, 2019.

\bibitem{liu_limited_1989}
D.~C. Liu and J.~Nocedal, ``\BIBforeignlanguage{en}{On the limited memory {BFGS} method for large scale optimization},'' \emph{\BIBforeignlanguage{en}{Math. Program.}}, vol.~45, no.~1, pp. 503--528, Aug. 1989.

\bibitem{ling_dealiasing_2023}
H.~J. Ling, O.~Bernard, N.~Ducros, and D.~Garcia, ``Phase {Unwrapping} of {Color} {Doppler} {Echocardiography} using {Deep} {Learning},'' \emph{IEEE Trans. Ultrason. Ferroelectr. Freq. Control}, vol.~70, no.~8, pp. 810--820, Aug. 2023.

\bibitem{chnafa_image-based_2014}
C.~Chnafa, S.~Mendez, and F.~Nicoud, ``Image-based large-eddy simulation in a realistic left heart,'' \emph{Comput. Fluids}, vol.~94, pp. 173--187, May 2014.

\bibitem{chnafa_image-based_2016}
C.~Chnafa, S.~Mendez, and F.~Nicoud, ``\BIBforeignlanguage{en}{Image-{Based} {Simulations} {Show} {Important} {Flow} {Fluctuations} in a {Normal} {Left} {Ventricle}: {What} {Could} be the {Implications}?}'' \emph{\BIBforeignlanguage{en}{Ann. Biomed. Eng.}}, vol.~44, no.~11, pp. 3346--3358, Nov. 2016.

\bibitem{faludi_left_2010}
R.~Faludi \emph{et~al.}, ``Left ventricular flow patterns in healthy subjects and patients with prosthetic mitral valves: {An} in vivo study using echocardiographic particle image velocimetry,'' \emph{J. Thorac. Cardiovasc. Surg.}, vol. 139, no.~6, pp. 1501--1510, Jun. 2010.

\bibitem{mehregan_doppler_2014}
F.~Mehregan \emph{et~al.}, ``Doppler vortography: a color {Doppler} approach for quantification of the intraventricular blood flow vortices,'' \emph{Ultrasound Med. Biol.}, vol.~40, no.~1, pp. 210--221, 2014.

\bibitem{ling_extraction_2023}
H.~J. Ling, N.~Painchaud, P.-Y. Courand, P.-M. Jodoin, D.~Garcia, and O.~Bernard, ``\BIBforeignlanguage{en}{Extraction of {Volumetric} {Indices} from {Echocardiography}: {Which} {Deep} {Learning} {Solution} for {Clinical} {Use}?}'' in \emph{\BIBforeignlanguage{en}{Proc. FIMH}}, 2023, pp. 245--254.

\bibitem{grossmann_can_2023}
T.~G. Grossmann, U.~J. Komorowska, J.~Latz, and C.-B. Schönlieb, ``Can {Physics}-{Informed} {Neural} {Networks} beat the {Finite} {Element} {Method}?'' Feb. 2023, arXiv:2302.04107 [math.NA].

\bibitem{sun_pipeline_2022}
Y.~Sun \emph{et~al.}, ``A {Pipeline} for the {Generation} of {Synthetic} {Cardiac} {Color} {Doppler},'' \emph{IEEE Trans. Ultrason. Ferroelectr. Freq. Control}, vol.~69, no.~3, pp. 932--941, Mar. 2022.

\bibitem{wang_preliminary_2021}
H.~Wang, S.~Gao, M.~Mozumi, M.~Omura, R.~Nagaoka, and H.~Hasegawa, ``Preliminary investigation on clutter filtering based on deep learning,'' \emph{Jpn. J. Appl. Phys.}, vol.~60, p. SDDE21, May 2021.

\bibitem{spoorthi_phasenet2.0_2020}
G.~E. Spoorthi, R.~K. Sai Subrahmanyam~Gorthi, and S.~Gorthi, ``Phasenet 2.0: Phase unwrapping of noisy data based on deep learning approach,'' \emph{IEEE Trans. Image Process.}, vol.~29, pp. 4862--4872, 2020.

\bibitem{nahas_deep_2020}
H.~Nahas, J.~S. Au, T.~Ishii, B.~Y.~S. Yiu, A.~J.~Y. Chee, and A.~C.~H. Yu, ``\BIBforeignlanguage{en}{A {Deep} {Learning} {Approach} to {Resolve} {Aliasing} {Artifacts} in {Ultrasound} {Color} {Flow} {Imaging}},'' \emph{\BIBforeignlanguage{en}{IEEE Trans. Ultrason. Ferroelectr. Freq. Control}}, vol.~67, no.~12, pp. 2615--2628, Dec. 2020.

\bibitem{krovetz_frequency_1974}
L.~J. Krovetz and S.~D. Goldbloom, ``Frequency {Content} of {Intravascular} and {Intracardiac} {Pressures} and {Their} {Time} {Derivatives},'' \emph{IEEE Trans. Biomed. Eng.}, vol. BME-21, no.~6, pp. 498--501, Nov. 1974.

\bibitem{posada_staggered_2016}
D.~Posada \emph{et~al.}, ``Staggered {Multiple}-{PRF} {Ultrafast} {Color} {Doppler},'' \emph{IEEE Trans. Med. Imaging}, vol.~35, no.~6, pp. 1510--1521, Jun. 2016.

\bibitem{ramalli_high-frame-rate_2020}
A.~Ramalli, A.~Rodriguez-Molares, J.~Avdal, J.~D’hooge, and L.~Løvstakken, ``High-{Frame}-{Rate} {Color} {Doppler} {Echocardiography}: {A} {Quantitative} {Comparison} of {Different} {Approaches},'' \emph{IEEE Trans. Ultrason. Ferroelectr. Freq. Control}, vol.~67, no.~5, pp. 923--933, May 2020.

\bibitem{ramalli_real-time_2018}
A.~Ramalli \emph{et~al.}, ``Real-{Time} {High}-{Frame}-{Rate} {Cardiac} {B}-{Mode} and {Tissue} {Doppler} {Imaging} {Based} on {Multiline} {Transmission} and {Multiline} {Acquisition},'' \emph{IEEE Trans. Ultrason. Ferroelectr. Freq. Control}, vol.~65, no.~11, pp. 2030--2041, Nov. 2018.

\end{thebibliography}

\end{document}